\documentclass[fleqn,usenatbib]{mnras}

\usepackage{newtxtext,newtxmath}

\usepackage[T1]{fontenc}
\usepackage{ae,aecompl}

\usepackage{fixmath}
\usepackage{hyperref}
\usepackage{booktabs}
\usepackage{multirow}
\usepackage{pdflscape}
\usepackage{dblfloatfix}
\usepackage{float}

\usepackage{graphicx}	
\usepackage{amsmath}	

\newcommand{\angs}{\textup{\AA}}

\newcommand{\kmsM}{\,km\,s$^{-1}$\,Mpc$^{-1}$}
\newcommand{\Ha}{\ion{H}{$\;\!\!\alpha$}}
\newcommand{\Hb}{\ion{H}{$\;\!\!\beta$}}
\newcommand{\Hii}{\ion{H}{II}}
\newcommand{\Oiii}{[\ion{O}{III}]}
\newcommand{\Nii}{[\ion{N}{II}]}
\newcommand{\mic}{\mu\mbox{m}}
\newcommand{\DMS}{$\mathrm{\Delta MS}$}
\newcommand{\DPS}{$\mathrm{\Delta PS}$}
\newcommand{\wi}{$\mathnormal{W1}$}
\newcommand{\wii}{$\mathnormal{W2}$}
\newcommand{\wiii}{$\mathnormal{W3}$}
\newcommand{\wiv}{$\mathnormal{W4}$}
\newcommand{\wxii}{$\mathnormal{W1-W2}$}
\newcommand{\wxxiii}{$\mathnormal{W2-W3}$}

\newcommand{\citeapos}[1]{\citeauthor{#1}'s (\citeyear{#1})}

\defcitealias{TSP20}{Paper~I}
\defcitealias{Kau+03}{K03}
\defcitealias{Kew+01}{K01}
\defcitealias{CidF+10}{C10}

\title[Multi-frequency characterization of activity in S0]{The local Universe in the era of large surveys - II.\ Multi-wavelength characterization of activity in nearby S0 galaxies}

\author[C.\ Jim\'enez-Palau et al.]{
C.\ Jim\'enez-Palau$^{1}$\thanks{E-mail: cjimenezp@icc.ub.edu, jm.solanes@ub.edu},
J.\ M.\ Solanes$^{1,2}$, J.\ D.\ Perea$^{3}$, A.\ del Olmo$^{3}$, and J.\ L.\ Tous$^{1,2}$
\\
$^{1}$Institut de Ci\`encies del Cosmos (ICCUB), Universitat de Barcelona, Mart\'i i Franqu\`es 1, E-08028 Barcelona, Spain\\
$^{2}$Departament de F\'isica Qu\`antica i Astrof\'isica, Universitat de Barcelona, Mart\'i i Franqu\`es 1, E-08028 Barcelona, Spain\\
$^{3}$Departamento de Astronom\'ia Extragal\'actica, Instituto de Astrof\'isica de Andaluc\'ia, IAA-CSIC, Glorieta de la Astronom\'ia s/n, E-18008 Granada, Spain
}

\date{Accepted: 2022 June 14. Revised: 2022 June 14. Received 2022 March 6}

\pubyear{2022}

\begin{document}
\label{firstpage}
\pagerange{\pageref{firstpage}--\pageref{lastpage}}
\maketitle

\begin{abstract}
This is the second paper in a series using data from tens of thousands S0 galaxies of the local Universe ($z\lesssim 0.1$) retrieved from the NASA-Sloan Atlas. It builds on the outcomes of the previous work, which introduced a new classification scheme for these objects based on the principal component analysis (PCA) of their optical spectrum and its projections on to the first two eigenvectors or principal components (the PC1--PC2 diagram). We provide a comprehensive characterization of the activity of present-day S0s throughout both the broad-band PC1--PC2 spectral classifier and the conventional narrow-line BPT/WHAN ones, contrasting the different types of activity classes they define, and present an alternative diagram that exploits the concordance between WHAN and PCA demarcations. The analysis is extended to the mid-infrared, radio and X-ray wavelengths by crossmatching our core sample with data from the \emph{WISE}, FIRST, \emph{XMM--Newton}, and \emph{Chandra} surveys. This has allowed us to carry out a thorough comparison of the most important activity diagnostics in the literature over different wavebands, discuss their similarities and differences, and explore the connections between them and with parameters related to star formation and black hole accretion. In particular, we find evidence that the bulk of nebular emission from radio and X-ray detected S0--Seyfert and LINER systems is not driven by star birth, while the dominant ionising radiation for a number of LINERs might come from post-AGB stars. These and other outcomes from the present work should be transferable to other morphologies.
\end{abstract}




\section{Introduction}
\label{S:intro}

\citet{Hub36} presented in his book {\it The Realm of Nebulae} a morphological classification scheme for galaxies based on their visual appearance. It included a hypothetical S0 type of objects showing traits halfway between those of the major lines of early-type galaxies (ETG), whose main representatives were the football-shaped ellipticals (E), and of late-type galaxies (LTG), containing different classes of spirals (S) that resembled flat, swirling discs. Shortly after, this S0 class was identified with the lens-shaped, armless disc galaxies frequently found in large aggregates of these objects. This identification was drawn on the fact that lenticular galaxies (henceforth, the terms lenticular and S0 will be used interchangeably) possess the general disc/bulge morphology of S galaxies, combined with a baryonic content dominated mainly by old stars and in which the warm and cold nebular components are usually scarce, as is the case in E galaxies.

The S0 is also the only morphological type that is relatively abundant in both low- and high-density environments, as shown by the morphology-density relation, first inferred by \citet{Dre80} for cluster regions and subsequently revised and extended to sparser environments by numerous authors \citetext{see e.g. \citealt{PG84}, \citealt{GHC86}, \citealt{Got+03}, \citealt{Cap+11}, and \citealt{Hou15}, among others}. Their dominance among the galaxies that populate the dense central regions of rich clusters led some to suggest that they were, in fact, the descendants of late-type progenitors whose gaseous interstellar medium (ISM) had been removed by direct collisions with other cluster members quenching their star formation \citep{SB51}. This mechanism was later superseded by the more plausible ram pressure produced by the hot intracluster medium (ICM) on the ISM of spiral galaxies as they move through it \citep{GG72}, probably accompanied by structural changes induced in the discs by the frequent, short-lived gravitational interactions they experience with the global cluster potential, with other companions, or with both, known as galaxy harassment \citep{Moo+96}.

Among the most important pieces of evidence supporting this scenario of S0s formation there is the already mentioned morphology-density relation, which convincingly demonstrates that in rich clusters the number fraction of S0s rises with local projected galaxy density at roughly the same pace that the fraction of S decreases. Furthermore, the nurturing effects of high-density environments on disc galaxies are also manifested over the time dimension in the form of a factor $\gtrsim 3$ increase in the fraction of lenticular objects in clusters between $z\simeq 0.4$ and $0$, accompanied by a commensurate decrease in the LTG fraction and an almost nil evolution of the abundance of ellipticals \citep[e.g.][]{Cou+94,Dre+97,Cou+98,Pog+99,Fas+00,Tre+03}.

Despite the body of evidence that substantiates the relevant role of clusters in the formation of S0, these galaxies are also found, albeit less frequently, in loose small groups and even in the general field -- with relative population fractions of $\sim 20$--30 per cent -- where the much thinner intergalactic medium and low peculiar velocities hinder the efficient functioning of hydrodynamic mechanisms. The presence of S0s in these sparser environments must then be explained by invoking close gravitational interactions between galaxies, in which at least the largest one is a LTG, that lead to their merger \citep[e.g.][]{Que+15}. The reason is that this formation mechanism reaches its peak efficiency when the relative speeds involved in the collision are comparable to the internal stellar motions of the colliding objects. This makes S0 special among the other Hubble types, as it suggests that galaxies from a single morphological class may have followed two radically different formation pathways.

The uniqueness of S0 galaxies has motivated the publication in recent years of a series of works, both numerical and observational, specifically dedicated to investigating these objects that have further reinforced this trait. Most of the experimental studies have focused on demonstrating, through controlled simulations with a wide variety of initial conditions, the feasibility of mergers (both major and minor) as a mechanism for the formation of galaxy remnants with structural and dynamic characteristics in accordance with those observed in the local population of S0, even in gas-poor implementations \citep[e.g.][]{Bor+14,MRM15,Que+15,EliM+18}. More recently, \citet{Dee+21} have used the publicly available data from the state-of-the-art IllustrisTNG-100 cosmological simulation \citep{Nel+19} to identify present-day S0 galaxies and investigate with unprecedented levels of detail their formation histories over cosmic time. This work shows that S0-like objects can, indeed, be produced by multiple mechanisms, with the main pathways of formation being by far merger events and the gas stripping that results from the infall of galaxy groups into larger units.

Observationally, the study of lenticular galaxies has begun to be approached through the analysis of large samples of optical spectra of these objects extracted either from single-fibre censuses or from the most recent surveys that use integral field spectroscopy. The former include research by \citet{Xia+16}, who have been among the first to analyse the local population of S0s and its relationship with evolutive processes from the point of view of activity classes. Based on the information provided by the Baldwin-Phillips-Terlevich (BPT) diagram \citetext{\citealt{BPT81}; see also section~\ref{SS:BPT}}, \citeauthor{Xia+16} were able to find evidence that activity and environment are closely related, noting that S0s with significant star formation and/or harbouring an AGN primarily reside in low-density regions \citep[see also][]{Dav+17}, while those showing a standard absorption-line spectrum or one with emission lines of low signal-to-noise ratio (SNR\;$<3$) are located in all kind of environments. More importantly, their results hinted at the existence of two types of S0s associated with different formation pathways. Although initially applied to the general galaxy population, the search of spectra with double-peak narrow emission lines by \citet{MM19} and \citet{Mas+20} can also be included within this category. These authors have found an important excess of S0s in samples of double-peak galaxies, accompanied by a systematic central excess of star formation. This and the fact that many of the double-peak lenticulars are isolated or located in poor groups has led them to suggest a scenario for these objects of bulge growth via multiple sequential minor mergers. For their part, spatially resolved spectroscopic measurements have also provided evidence of the existence of a subpopulation of field S0 galaxies that, compared to their classical counterparts located preferentially in groups and clusters, are less massive and show lower and flatter velocity dispersion profiles \citep{FraM+18,DomS+20}, as well as less rotational support \citep{Coc+20,Dee+20,Xu+22}.

The duality in the properties of lenticular galaxies has been further confirmed by the extensive statistical study of a massive database of more than $68,\!000$ single-fibre optical spectra of S0 galaxies in the local Universe ($z\lesssim 0.1$) led out recently by some of us \citep[][hereafter \citetalias{TSP20}]{TSP20}. In that work, we applied the principal component analysis (PCA) technique to reduce the large number of dimensions of the spectral data to the features encoded in the low-dimensional space defined by the first, most relevant eigenspectra (see Appendix~\ref{A:eigenspectra}), while minimizing information loss. The projections of the S0 spectra on the axes of the bivariate subspace defined by the first two eigenvectors or principal components, denoted here by PC1 and PC2, explains about 90 per cent of the total variance and reveals the existence of two main regions outlined by subpopulations of lenticular galaxies with statistically inconsistent physical properties. We call these regions `Passive Sequence' (PS) and `Active Cloud' (AC), since they encompass sources  whose spectra are representative, respectively,  of passive and active galaxies. Compared to their absorption-dominated counterparts, S0s with significant nebular emission are, on average, somewhat less massive, more luminous with less concentrated light profiles, have a younger, bluer and metal-poorer stellar component, and avoid high-galaxy-density environments. A narrow dividing zone or `Transition Region' (TR), formed by objects with intermediate spectral characteristics, separates the two main areas of the PC1--PC2 subspace and completes a novel classification for the S0s that is reminiscent of the well-known `Red Sequence-Green Valley-Blue Cloud' division applied to the entire galaxy population in colour-magnitude diagrams. Interestingly, the analysis in \citetalias{TSP20} also revealed that most of the S0s included in the AC class, which account for at least a quarter of the population of nearby lenticular galaxies, have star formation rates (SFRs) fully consistent with those observed in late spirals. This is in line with the findings of \citet{Kav+07} from near-ultraviolet data. These results therefore indicate that present-day S0s with abundant star formation are not an isolated phenomenon and, consequently, that the traditional conception of these galaxies as basically red and passive stellar systems should be replaced by that of a class of objects that span a range of physical properties, can exhibit different levels of activity and, so it seems, also follow diverse formation channels \citep[see also][]{WS03, Mor+06,Cro+11,Bar+13}.

The present work aims to further deepen our understanding of the properties of the S0 population in the low-$z$ Universe by carrying out an exhaustive characterization of the activity of these objects, as much linked to black hole accretion as to normal stellar processes. This effort is particularly relevant given the aforementioned confirmation of the existence of a significant fraction of lenticular galaxies with ongoing star formation. This fact, together with the higher fractional abundance of nuclear activity in galaxies with a strong bulge component \citep[e.g.][]{Ove+03,Ho08} and the tendency of systems hosting an active central supermassive black hole (SMBH) to have global SFRs typical of pure star-forming galaxies \citep{Ros+13,Suh+19}, make S0s ideal systems for ascertaining the extent to which nuclear activity and star formation are interrelated phenomena. Besides, both observations and theory suggest that there is a link between the activity of galaxies and the evolution of the global properties of their hosts, the similarity between the space density of quasars and the integrated star formation history of the Universe being a good example \citetext{see e.g. \citealt{Che+22} and references therein}. Therefore, a study of these characteristics can provide valuable information, also for the general population of galaxies, for a better understanding of the physical nature of this connection, as well as helping to determine whether it is direct or indirect. 

This article is organized as follows. We begin by describing in section~\ref{S:optical} the contents of the baseline data set used to carry out this endeavour. Section~\ref{S:spectral_diagnostics} is devoted to a thorough comparative statistical study of the various classes of activity into which lenticular galaxies can be subdivided according to the information contained in their optical spectra. We start by describing the recently introduced PCA-based spectral classification (section~\ref{SS:PCA}) and continue with the conventional BPT and WHAN schemes (sections~\ref{SS:BPT} and \ref{SS:WHAN}). In section~\ref{SS:DPSN} a fourth alternative diagnostic of activity is introduced that uses elements from both the PCA and WHAN classifiers. Limiting the investigation of activity in galaxies to only the optical window can lead to incomplete and biased results due to obscuration by dust or mixed emission from different sources \citep[see e.g.][]{Sat+08,Sat+14,Suh+19}. For this reason, we have extended our analysis to other wavebands by matching our optical data on S0s to mid-infrared (mid-IR) extragalactic sources (section~\ref{S:midIR}), as well as to radio and X-ray measurements (section~\ref{S:radio_X-ray}). Finally, we summarise our main findings in section~\ref{S:summary}, while the basic ideas behind our PCA treatment of the S0 spectral data are outlined in Appendix~\ref{A:eigenspectra}. Given the breadth of this study, it has been considered convenient to postpone the discussion on the implications of the results obtained regarding the possible evolutionary scenarios of lenticular galaxies to the forthcoming papers in this series. 

When necessary, it is assumed that we live in a standard concordant $\mathrm{\Lambda}$CDM universe with cosmological parameters: $\mathrm{\Omega_m} = 0.3$, $\mathrm{\Omega_\Lambda} = 0.7$ and $H_0 = 70$ \kmsM.

\section{Baseline Data on Present-day S0 galaxies}
\label{S:optical}

Our core database is an extensive share of the original sample used in \citetalias{TSP20}. It consists of a compilation of spectrophotometric measurements for $56,\!008$ S0 galaxies having elliptical $r$-band Petrosian magnitudes $\leq 17.77$, heliocentric redshifts $0.01\leq z\leq 0.1$ and good-quality optical spectra, extracted from the reprocessed Sloan Digital Sky Survey \citetext{SDSS; \citealt{Yor+00}} photometry listed in the \texttt{v1\_0\_1} version of the NASA-Sloan Atlas\footnote{This data set is built around the Main Galaxy Sample \citetext{\citealt{Str+02}} of the SDSS, an essentially complete magnitude-limited subset of the Legacy Survey.} \citep[NSA;][]{Bla+11} and other public catalogs. SDSS measurements are complemented in the NSA with observations from the CfA Redshift Catalog \citep[ZCAT;][]{HGC95}, the 2dF Galaxy Survey \citep[2dFGRS;][]{Col+01}, the 6dF Galaxy Survey \citep[6dFGS;][]{Jon+04} and the NASA/IPAC Extragalactic Database (NED), with the addition of ultraviolet data from the \textit{Galaxy Evolution Explorer} \citep[\textit{GALEX};][]{Bos+11} satellite and a small fraction of 21-cm measurements of the Arecibo Legacy Fast ALFA \citep[ALFALFA;][]{Gio+05} survey in the overlapping regions. Among other properties, our database includes two numerical parameters derived by \citet{DomS+18} from deep learning models: the revised Hubble-type index, $T$ \citep{dVau77}, and the probability that an ETG is an S0, $P_{\mathrm{S0}}$, which we use to robustly identify galaxies of S0 morphology by imposing the conditions $T\leq 0$ and a $P_{\mathrm{S0}} > 0.7$. For the stellar masses, $M_\ast$, we prefer the estimates listed in the GALEX-SDSS-WISE Legacy Catalog 2 \citep[GSWLC-2;][]{SBL18}, which also acts as the source for the SFRs used in this work, while fluxes and equivalent widths, EW, of the most important nebular emission lines, together with the BPT spectral classes (see section~\ref{SS:BPT}) are retrieved from the Portsmouth stellar kinematics and emission-line flux measurements by \citet{Tho+13}. All this information is completed with several parameters calculated in \citetalias{TSP20}, such as two estimators of the local density, one based on the average distance to the first fifth neighbours in a volume-limited subset of our core database formed by all galaxies with $M_r < -20.5$, and one that ranks this same type of estimator according to the percentile that each galaxy gets in the deepest volume-limited subset that can be defined for it within this database \citetext{see \citealt{Tou18} for details}. Besides, we have the coefficients of the projections of the entire optical spectra of S0 galaxies into the first principal components (see also Appendix~\ref{A:eigenspectra}), as well as their PCA-based activity classification (see section~\ref{SS:PCA}) inferred from the spectral analysis of this population carried out in \citetalias{TSP20}.

\section{Classification of Present-day S0 galaxies from their Optical Spectra}
\label{S:spectral_diagnostics}

In this section, we perform a comparative statistical analysis of the different classes of activity into which the population of nearby S0 galaxies can be subdivided by applying various treatments to the information contained in their optical spectra.

\subsection{Spectral classes from the PCA of the full spectrum}
\label{SS:PCA}

\begin{figure}
	\includegraphics[width=\columnwidth, viewport=10 -85 420 315, clip=true]{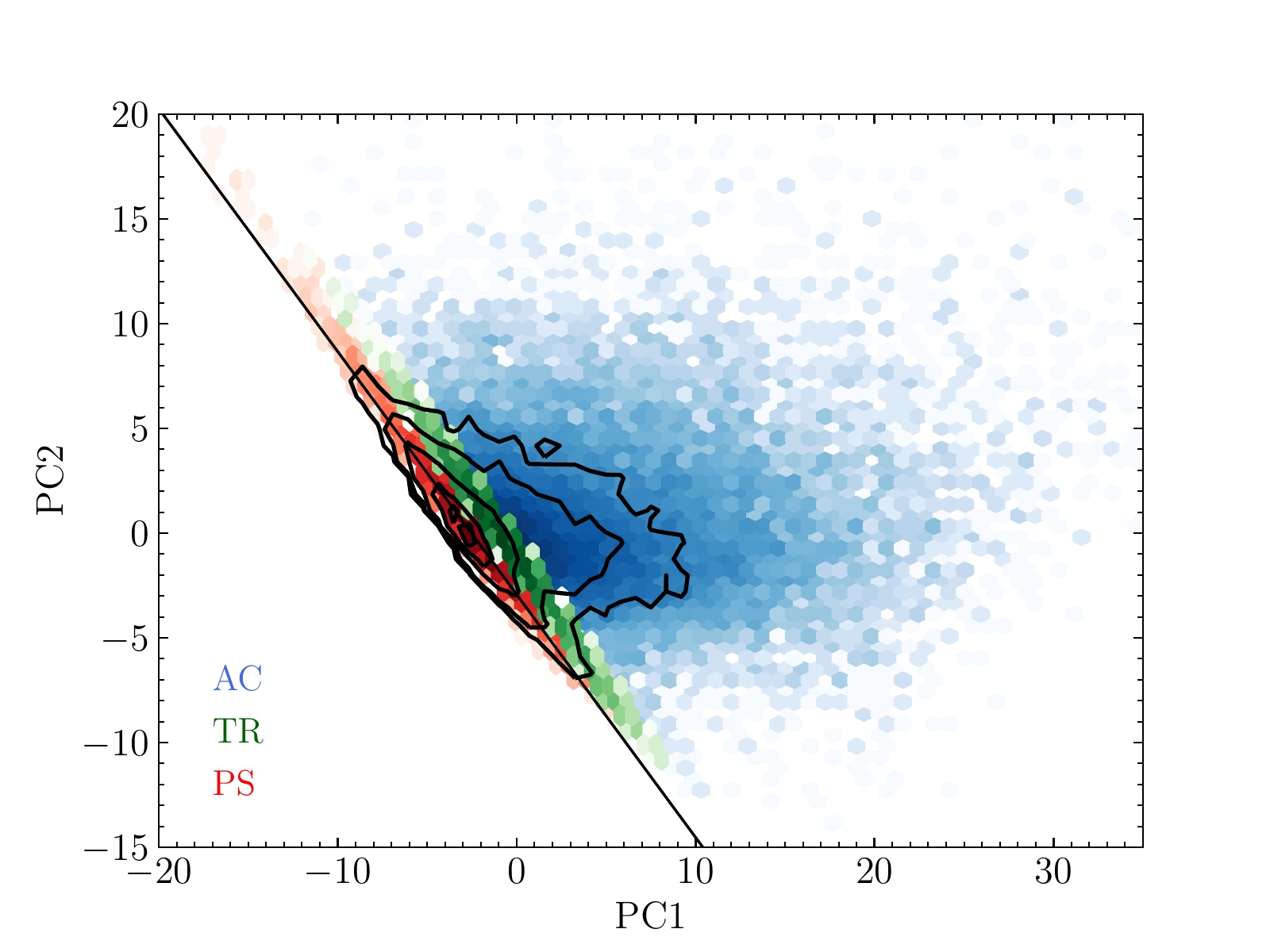}
	\vspace*{-6\baselineskip}
    \caption{Projections of the optical spectra of our core sample of local S0 galaxies on their first two Principal Components defined in \citetalias{TSP20}. The PC1--PC2 diagram is subdivided into three distinct regions corresponding to three different spectral classes throughout heat maps of different colours: PS (red), TR (green) and AC (blue). The intensity of all colours scales with the logarithm of the number density of points which are binned in hexagonal cells. The overlaid black contours correspond to global number densities 20, 40, 60, 80, and 90 per cent of the peak value drawn from a volume-limited subset of S0s taken from the full sample that includes galaxies with a Petrosian $r$-band absolute magnitude $M_r < -20.5$. The diagonal black straight line shows the PS ridge.}
    \label{F:pc1-pc2_dens}
\end{figure}

\begin{figure}
	\includegraphics[width=\columnwidth, viewport=1 1 440 620, clip=true]{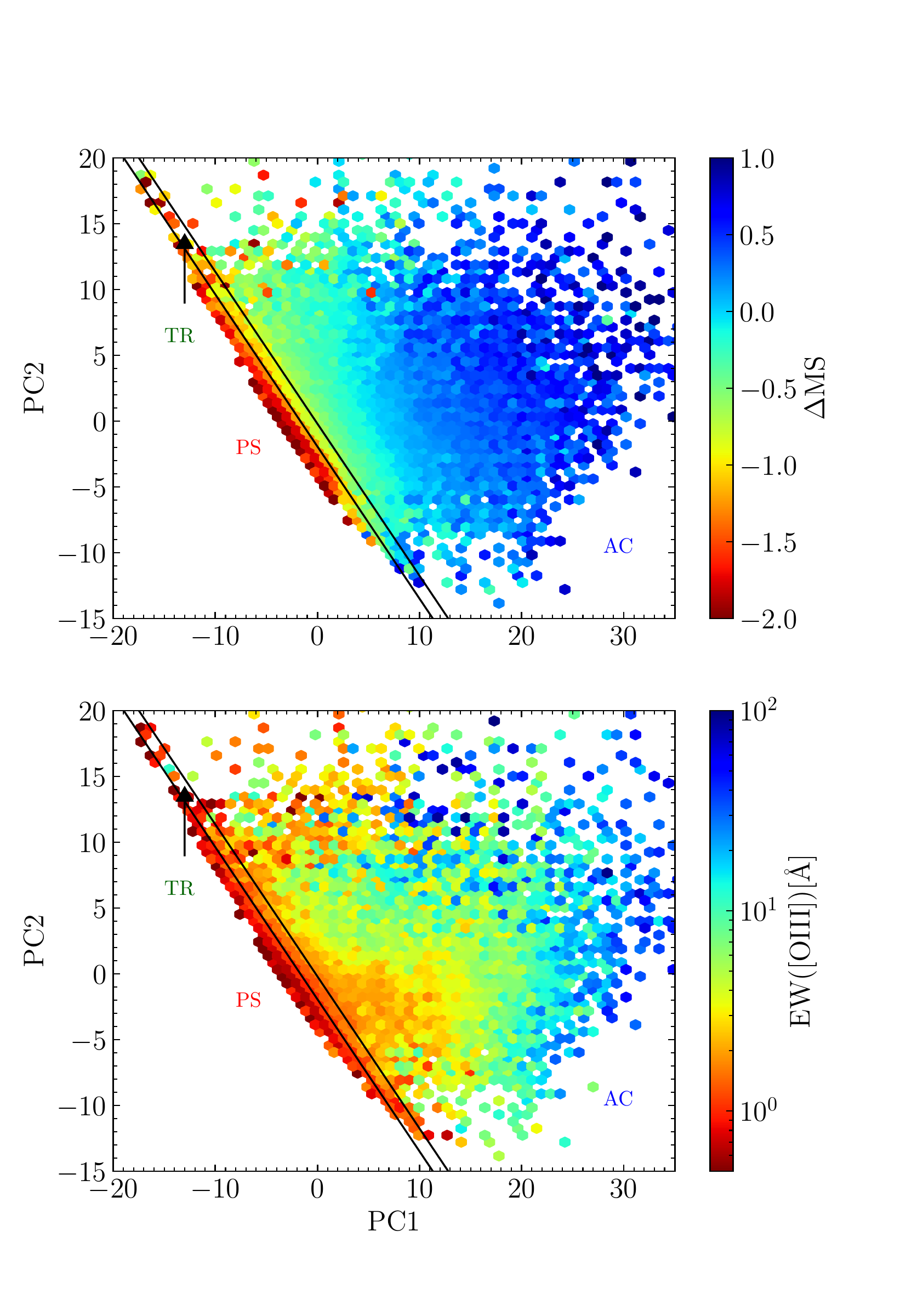}
	\vspace*{-2.5\baselineskip}
    \caption{Alternative versions of the PC1--PC2 diagram for our core sample of nearby S0 galaxies. The scatter plots include a colour code that varies according to the mean value of \DMS\ (eq.~(\ref{E:DMS}); \emph{top}) and the \Oiii$\lambda 5007\,\angs$ EW (\emph{bottom}) in each of the hexagonal cells in which the data has been grouped. Two parallel black straight lines show the best linear approximations to the PS/TR and TR/AC dividers. Note that the number of measurements used to build both plots is not exactly the same, since the data in the bottom panel obey the additional constraint $\mbox{AoN(\Oiii)} > 1.5$ (see text).}
    \label{F:pc1-pc2_DMSO3}
\end{figure}

As commented in the Introduction and shown in Fig.~\ref{F:pc1-pc2_dens}, the projections of the optical spectra of local S0 galaxies on their first two principal components enable their classification into three distinct spectral classes: PS, formed by the members of the Passive Sequence, the compact and densely populated narrow band that diagonally crosses the PC1--PC2 subspace (in red colours); AC, which comprises the galaxies located in the substantially less populated and much more scattered Active Cloud zone running from the right of the PS (blue); and TR, which contains the objects located in the slim Transition Region that separates the PS and AC areas (green). It can also be seen from Fig.~\ref{F:pc1-pc2_dens} that the definition of the PCA spectral groups correlates very strongly with the projected distance of galaxies to the ridge line of the PS in the PC1--PC2 plane, which obeys the equation
\begin{equation}
    \mbox{PC2}=-1.162\cdot\mbox{PC1}-2.932\;.
    \label{E:PSridge}
\end{equation}
This distance, dubbed \DPS, can be inferred from the function:
\begin{equation}
    \mathrm{\Delta PS}=\log\left[2.913+0.758\cdot\mbox{PC1}+0.652\cdot\mbox{PC2}\right]\;,
    \label{E:DPS}
\end{equation}
whose dominion encompasses all physically plausible values of PC1 and PC2, which are essentially those in the half-plane to the right of the ridge line in Fig.~\ref{F:pc1-pc2_dens} (see \citetalias{TSP20}), and that reduces the PCA taxonomy to a one-dimensional classifier of activity. Much as with the traditional BPT and WHAN diagrams (see next sections), the spatial distribution shown by the broad-band spectral classes of S0 galaxies over this subspace provides a fairly faithful representation of the region spanned by the optical spectra of all Hubble types (see fig.~11 in \citetalias{TSP20}), although in terms of occurrence most of the lenticulars are concentrated in the PS.

Fig.~\ref{F:pc1-pc2_DMSO3} shows two alternative versions of the PC1–PC2 diagram in which a colour scale is added to encode activity information. The extra dimension \DMS\ in the top panel indicates the average distance to what is often referred to as the galaxies’ Main Sequence (MS). This is a linear relationship in a log-log plot \citep[but see][]{Pop+19} between the SFR and $M_\ast$ obeyed by most present-day star-forming galaxies, used here to measure the level of activity shown by nearby S0s associated with the formation of new stars. This ‘distance’ is inferred through the difference between the observed SFR and that expected from a MS counterpart with the same $M_\ast$, via the equation
\begin{equation}
\mathrm{\Delta MS} = \log\left[\frac{\mathrm{SFR}}{\mbox{M}_\odot\,\mathrm{yr}^{-1}}\right]-0.76\log\left[\frac{M_\ast}{\mbox{M}_\odot}\right] + 7.6\;,
\label{E:DMS}
\end{equation}  
where the expression providing the location of the MS ridge line for local objects has been taken from \citet{RP15}. Note that due to the explicit dependence on $M_\ast$, this equation actually informs about variations in the specific SFR (SSFR).

Due to the lack of an equivalent way to isolate the contribution of the central SMBH to the global activity of galaxies, we have chosen to use in the lower panel of Fig.~\ref{F:pc1-pc2_DMSO3} the EW of the $\Oiii\lambda 5007\,\angs$ emission line to encode the AGN-related activity. This is a spectral line excited primarily by electron collisions that is generally the strongest and least blended of the emission lines originated in the Narrow Line Region (NLR) of AGN-dominated galaxies, and that is often assumed to provide an unbiased measure of the ionising flux from the nucleus \citetext{e.g.\ \citealt{BL05,Li+06,Com+09,KH09}; see also \citealt{TB10} and references therein}. Besides, the use of an EW measure makes it unnecessary to correct for NLR dust extinction. Note also that in order to provide clean maps we only show, in this and the remaining figures in this work involving spectral diagnostic diagrams, those galaxies with a good-quality optical spectrum, as defined in \citetalias{TSP20}, and for which the emission lines used in the diagrams have values of the amplitude-over-noise parameter, AoN, greater than $1.5$, which is the minimum condition for a reliable estimation of the line flux according to the Portsmouth catalog. 

\begin{table*}
	\centering
	\caption{Demarcations of the spectral classes that can be defined in BPT--NII diagnostic diagrams.}
	\label{T:demarcations}
	\begin{tabular}{cccc} 
	\hline
	Activity classes & Equation of divider & Reference & Acronym\\
	\hline
	Star-forming/Composite & $\log (\Oiii/\Hb)\,=\,0.61/(\log(\Nii/\Ha)-0.05)+1.30$ & \citet{Kau+03} & \citetalias{Kau+03}\\
    Composite/AGN & $\log(\Oiii/\Hb)\,=\,0.61/(\log(\Nii/\Ha)-0.47)+1.19$ & \citet{Kew+01} & \citetalias{Kew+01}\\
    Seyfert/LINER & $\log(\Oiii/\Hb)\,=\,1.01\log(\Nii/\Ha)+0.48$ & \citet{CidF+10} & \citetalias{CidF+10}\\
	\hline
	\end{tabular}
\end{table*}

\begin{figure*}
	\includegraphics[width=\textwidth, viewport=60 1 840 530, clip=true]{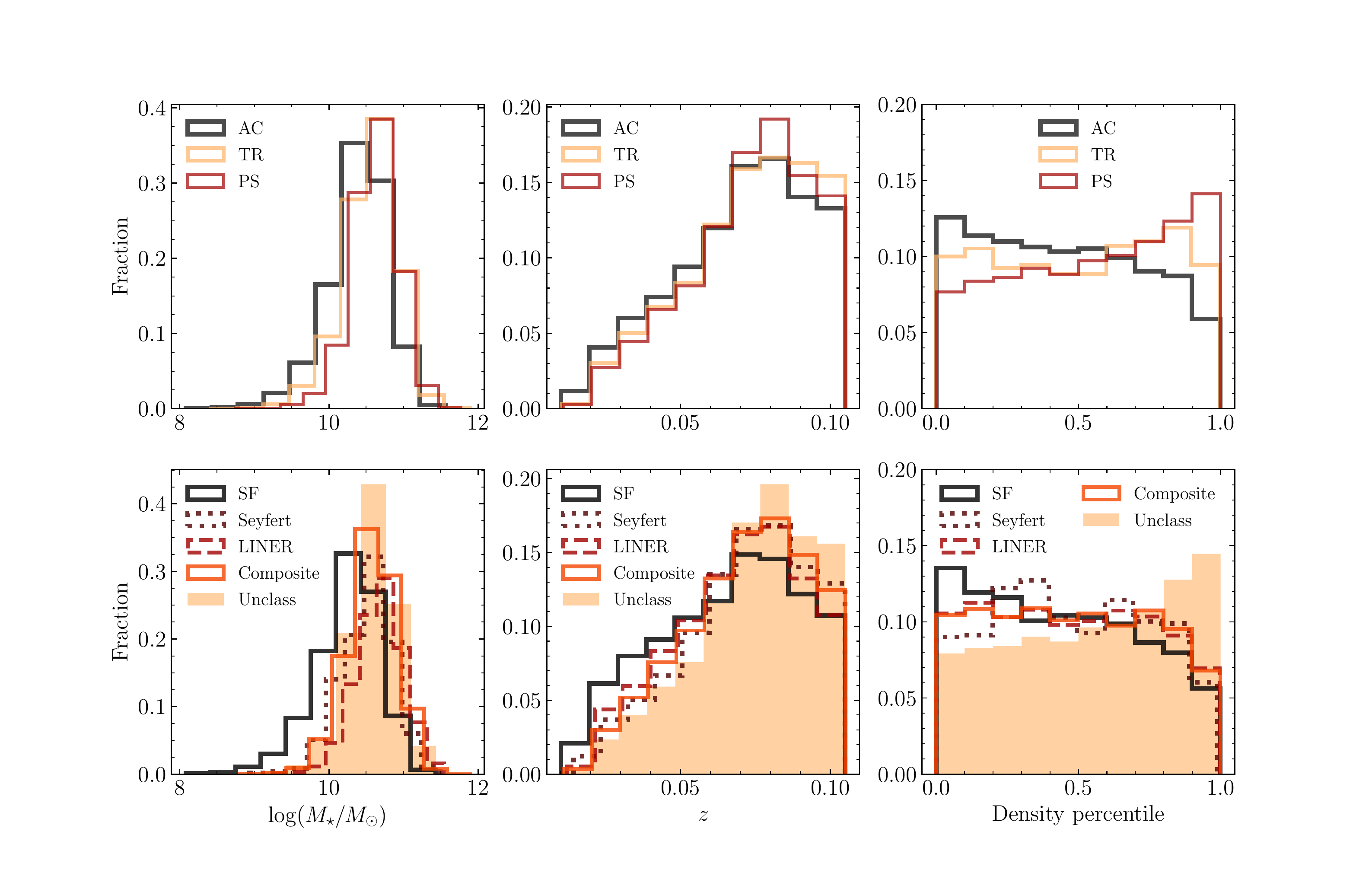}
	\vspace*{-2\baselineskip}
    \caption{Distributions of stellar mass (\emph{left}), heliocentric redshift (\emph{middle}) and percentile rank of the density (\emph{right}) for our sample of nearby S0 galaxies in terms of both the broad-band PCA (\emph{top}) and narrow-line BPT (\emph{bottom}) spectral classes. To facilitate intercomparison among these distributions, the total counts in all subsets have been normalized to sum to one, so the height of the histograms represents the proportion of a given class in each one of the bins defined on the $x$-axes.}
    \label{F:histo_bpt_pca}
\end{figure*}

Comparison of the two panels of Fig.~\ref{F:pc1-pc2_DMSO3} shows that the lowest values of both \DMS\ and EW(\Oiii) are achieved around the PS region, which indicates that this spectral class mainly encompasses galaxies with null or depressed star formation and probably lacking nuclear activity. In contrast, the highest levels of activity normally occur in galaxies that inhabit the section of the AC region furthest from the PS. Nevertheless, the parameter \DMS\ seems to be somewhat better correlated with the distance to such sequence (the colour scale runs more parallel to the PS ridge) than the EW(\Oiii).

In the top three panels of Fig.~\ref{F:histo_bpt_pca}, we depict the distributions of the PCA spectral classes -- conveniently renormalized, so they can be compared on an equal basis -- in terms of what is probably the most fundamental global physical property of galaxies, $M_\ast$, and of two extrinsic parameters, $z$ and a measure of the influence of environment given by the percentile rank associated with our nearest-neighbour-based local density estimator. As shown by these graphs, lenticular AC galaxies tend to be somewhat less massive than those belonging to the PS and TR classes, which exhibit fairly similar stellar mass distributions to each other. Likewise, the redshift distribution of this spectral class shows a mild bias towards low redshifts compared to those from the other two groups. For their part, the distributions of the PCA types according to our local environment descriptor reveal significant differences, with the S0 members of the PS and AC classes showing opposite tendencies, the latter being more abundant in regions devoid of galaxies and the former in the most crowded environments \citetext{see also fig.\ 10 in \citetalias{TSP20}}. In contrast, the distribution of the lenticulars of the TR class shows a roughly neutral behaviour with local density. Our data set contains $32,\!519$, 4259 and $19,\!230$ lenticular galaxies that are, respectively, bona fide members of the classes PS, TR and AC. So, not surprisingly, most of the nearby S0s are objects with a passive optical spectrum, although the number of their counterparts with signs of mild or substantial activity is by no means negligible.

\subsection{Spectral classes from ratios of narrow emission lines}
\label{SS:BPT}

\begin{figure}
	\includegraphics[width=\columnwidth, viewport=0 0 350 600, clip=true]{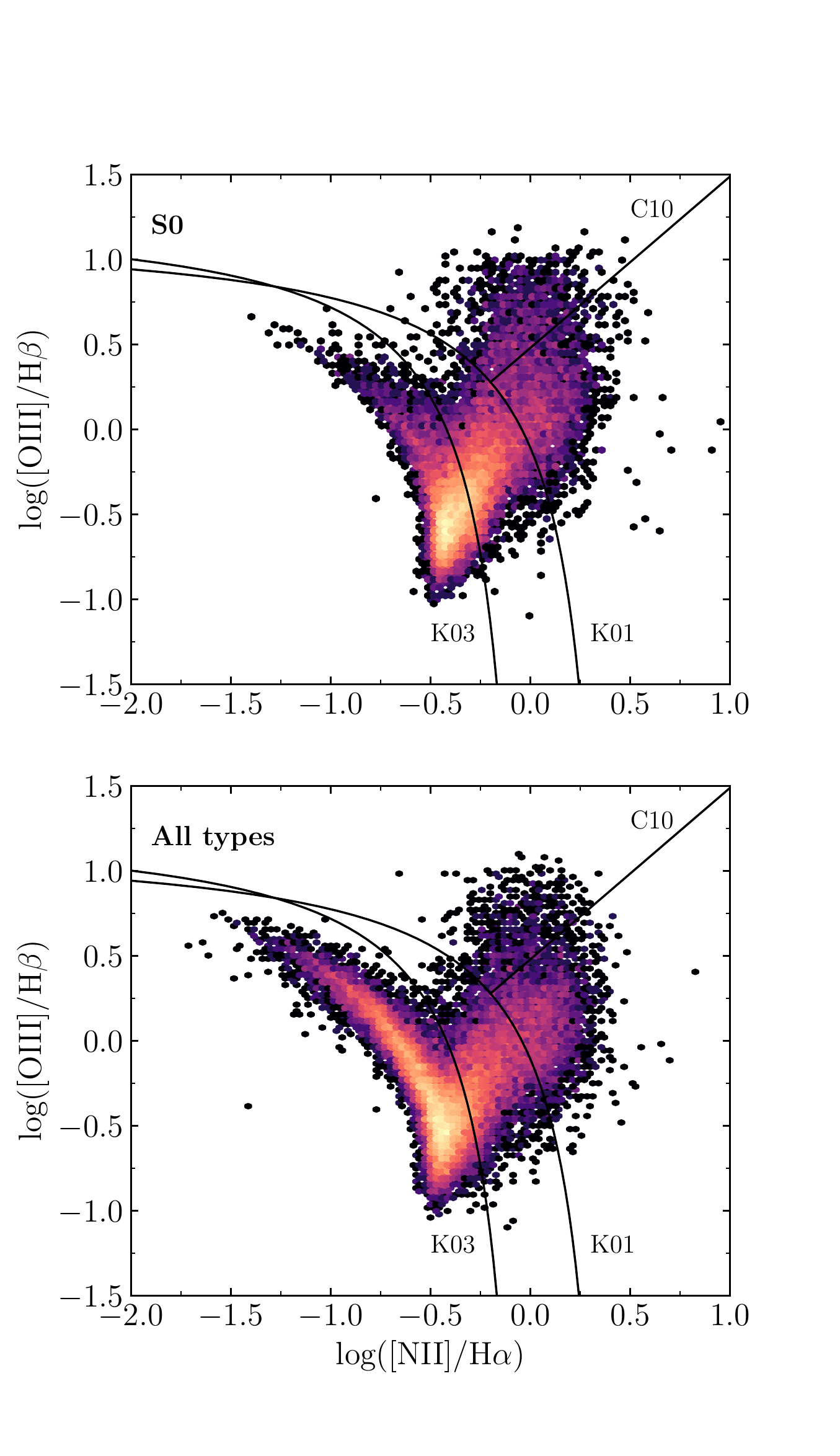}
	\vspace*{-3.0\baselineskip}
    \caption{Heatmaps of the BPT--NII diagrams for our sample of nearby S0 galaxies (\emph{top}) and for an arbitrary sample of $\sim 45,\!000$ galaxies of all Hubble types with $z\leq 0.1$ (\emph{bottom}). The latter has been assembled from $\sim 36,\!000$ galaxies of types E, Sa, Sb, and Sc, plus about 9000 S0 galaxies randomly removed from the upper panel to guarantee that the figure compares two fully independent data sets. Only galaxies from the baseline data set having both a good-quality optical spectrum and values $\mbox{AoN} > 1.5$ for all four emission lines have been included in the diagrams. In both panels, colours scale with the logarithm of the number of points in each hexagonal cell, with darker colours corresponding to smaller values and lighter colours to larger ones. Black curves show the demarcation lines listed in Table~\ref{T:demarcations}.}
    \label{F:BPT_dens_S0-All}
\end{figure}

The BPT diagnostic diagrams are a well-known set of plots comparing pairs of narrow optical emission-line ratios designed to distinguish the ionisation mechanism of the gaseous component. They have become one of the benchmarks for the identification of the different classes of activity in galaxies. In Fig.~\ref{F:BPT_dens_S0-All}, we show heatmaps of the most extensively used version of this classification scheme depicting the $\Nii\lambda 6584/\Ha$ flux ratio against the $\Oiii\lambda 5007/\Hb$ flux ratio \citetext{the BPT--NII diagram; see fig.\ 5 of \citealt{BPT81}} for our sample of lenticular objects (top panel) and for an arbitrary sample of similar size of nearby galaxies extracted from the SDSS having an optical spectrum of good quality (see \citetalias{TSP20}) and with morphologies evenly distributed in the range of Hubble types from E to Sc (bottom panel). The black curves in the diagrams mark the boundaries of the regions traditionally adopted to distinguish among the different activity classes: star-forming galaxies (SF), Seyfert galaxies, which host a considerable active galactic nucleus (AGN), low-ionisation nuclear emission-line region galaxies (LINER), encompassing true low-luminosity AGNs as well as galaxies whose LINER-like emission is inconsistent with ionisation by a nuclear source, and Composite sources that show mixed SF and AGN contributions\footnote{The Composite class is considered ill-defined by some authors, who therefore exclude it from their emission-line taxonomic paradigms of galaxies \citep[e.g.][]{CidF+11}. See also section~\ref{SS:WHAN}.}. The demarcations, summarised in Table~\ref{T:demarcations}, are the same used in the Portsmouth catalog, which is our reference source for the BPT classification. Comparison of both panels shows that the two galaxy samples define fairly similar `seagull-shaped' loci in BPT--NII space, the biggest difference being a somewhat emptier upper half of the `left wing' on the part of the S0, due to a lower presence of objects with intense star formation. Consequently, it is reasonable to conclude that, except for minor details, the spatial distribution of present-day lenticular galaxies in BPT--NII diagnostic diagrams bears a resemblance to that of the entire Hubble Sequence.

\begin{figure}
	\includegraphics[width=\columnwidth, viewport=1 1 460 650, clip=true]{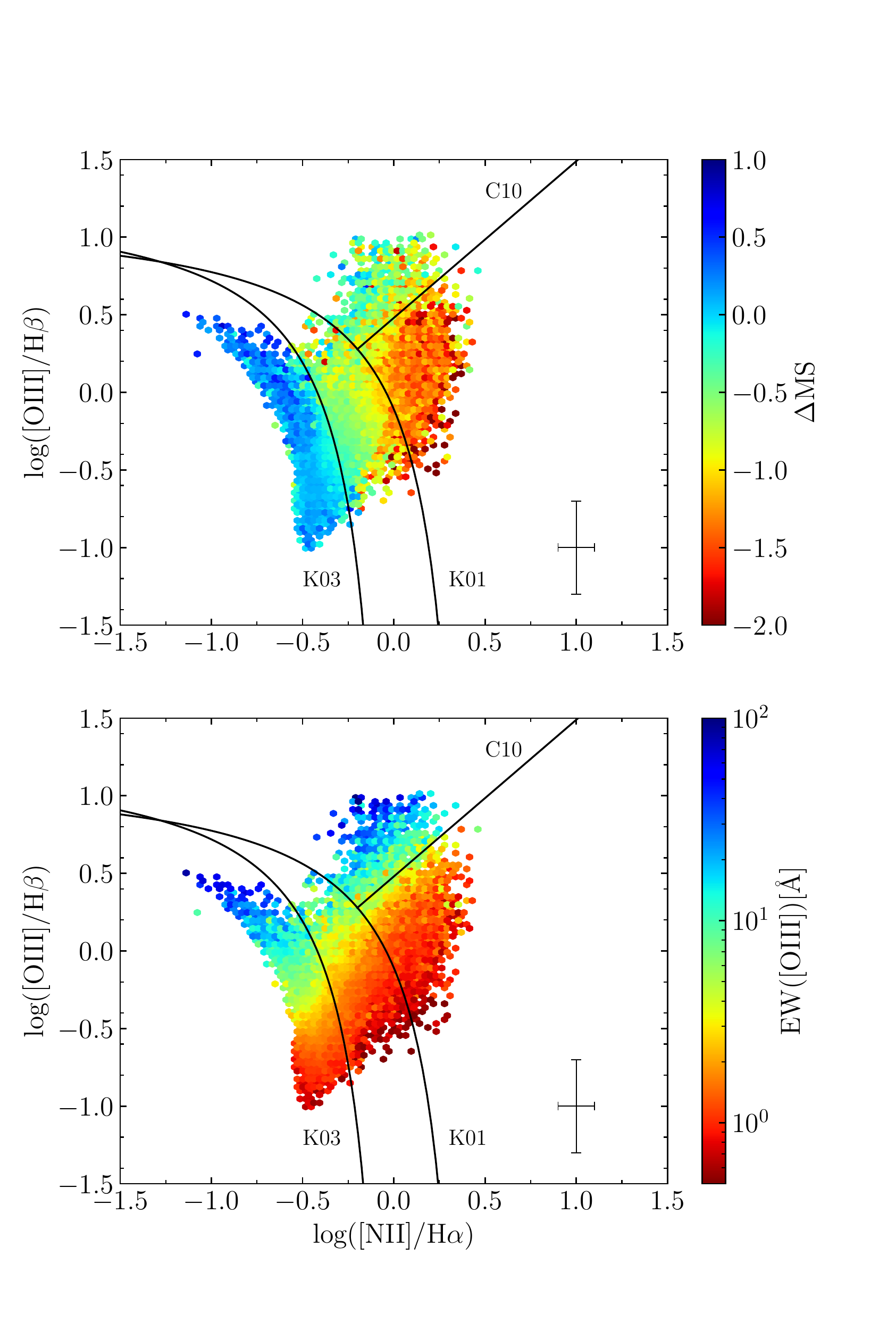}
	\vspace*{-2.5\baselineskip}
    \caption{Same as in Fig.~\ref{F:pc1-pc2_DMSO3} but for the BPT--NII diagram. Black curves show the demarcation lines listed in Table~\ref{T:demarcations}. As in Fig.~\ref{F:BPT_dens_S0-All}, we only plot galaxies whose parameters pass our quality filters. The error bars in the bottom-right corners of the panels show the medians of the errors in both axes.}
    \label{F:bpt_DMSO3}
\end{figure}

In Fig.~\ref{F:bpt_DMSO3}, colour bars are added to the BPT--NII diagram to report, as in Fig.~\ref{F:pc1-pc2_DMSO3}, on the mean values of the distance to the galaxies’ MS, \DMS\ (top panel), and of the EW(\Oiii) (bottom panel). As would be expected, these two representations of the BPT--NII diagram show that the highest values of this parameter essentially populate the left-wing sequence associated with star-forming galaxies, whose right-upper boundary is very well outlined by the \citetalias{Kau+03} demarcation. Although to a lesser extent, S0s with similarly high values of \DMS\ are also present in the left half of the right wing, extending from the bottom of the diagram to the Seyfert's region (some blue data points can be seen protruding from the wing tip). This area of the BPT--NII diagram, however, brings together galaxies with vastly different SSFRs, which makes the \DMS\ parameter show intermediate typical values across it. This dilution effect also affects the galaxies with the most depressed star-formation activity which, far from being restricted to the LINER region, also populate the lower part of the diagram. Such behaviours can be guessed by comparing these diagrams with the corresponding ones in the following section. For its part, the bottom panel shows that the \Oiii\ EW decreases in a more or less orderly fashion towards the bottom-right of the diagram, following a direction nearly perpendicular to the \citetalias{CidF+10} divisor. The fact that the highest values of this emission line are located at both the Seyfert and SF wingtips demonstrates that star-forming galaxies can also be strong \Oiii\ emitters \citep[e.g.][]{Suz+16}.

The distributions of the total stellar mass, redshift and density percentile for the BPT spectral classes are depicted in the bottom panels of Fig.~\ref{F:histo_bpt_pca}. They are drawn from a total of $9490$ SF lenticulars, $1171$ Seyfert, $3678$ LINER, and $8459$ Composite. A fifth `Unclass' type with $33,\!210$ members has been added to this scheme to account for objects with either an unreliable BPT classification or none at all because they have one or more emission lines with a low contrast (i.e.\ small AoN) or a total lack thereof (see Table~\ref{T:BPT-PCA_vis}). As the first two panels show, the distributions of $M_\ast$ and $z$ behave quite similarly in most cases, the only exception being the SF class, whose members tend to be less massive and exhibit redshifts skewed towards lower values compared to those of the other spectral groups. This disparate behaviour is simply a present-epoch manifestation of downsizing -- also latent in the histograms of the PCA classes since, as Table~\ref{T:BPT-PCA_vis} shows, essentially all SF lenticulars (more than 99 per cent of this BPT group) are members of the AC class -- combined with the fact that the different subsets of S0 galaxies behave largely as flux-limited samples \citep[see e.g.\ the upper panel of fig.\ 1 from][]{Gki+21}. This finding suggests that the signature of downsizing in the local Universe could also be preserved among individual Hubble types. Regarding the relationship with the local environment, Seyfert, LINER and Composite S0s display relatively uniform abundances for densities below the 90th percentile and an appreciable drop in the relative fractions in the last 10 per cent, while the SF lenticulars show a monotonous decrease in their abundance with increasing density percentile. Only the sources labelled Unclass, which are mainly passive galaxies with absorption spectra (see below), clearly favour high-density environments. 

\begin{table}
\centering
\caption{Joint distribution of the PCA and BPT spectral classes for S0 galaxies with $z\leq 0.1$.}
\label{T:BPT-PCA_vis}
\begin{tabular}{cccc} 
\hline
		BPT class &  & PCA type& \\
		 & Active Cloud & Transition Region & Passive Sequence \\
		\hline
Star-forming & 9414&53&23  \\
Seyfert &1073&75&23\\
LINER & 631&625&2422 \\
Composite &4971&1691&1797\\ 
Unclass$^a$ & 3141&1815&28254\\
\hline
\multicolumn{4}{l}{\footnotesize($^a$) Encompasses galaxies with unreliable or absent BPT classification.}
\end{tabular}
\end{table}

To complete this benchmarking exercise for the optical PCA and BPT spectral types of the S0 galaxies in the local Universe, we show in Table~\ref{T:BPT-PCA_vis} their joint distribution. Practically all SF and Seyfert lenticulars belong to the AC, which also contains a substantial fraction of galaxies classified as Composite ($\sim 30$ per cent of the objects with a BPT tag). In contrast, the PS class is fully dominated by galaxies that lack a BPT classification (about 60 per cent of all the objects in our baseline S0 sample), showing that these sources can nevertheless be perfectly characterised by means of the PCA taxonomy. Most of the S0 LINERs are also members of the PS class, so while a good handful of them are found in the TR and AC regions ($\sim 17$ per cent each), it is clear that this class of galaxies sustains, in general, a low level of activity, of whatever type (see sections~\ref{SS:WHAN} and \ref{SS:radio_X-ray_nature}). Furthermore, S0s included in the Composite class live up to their name and are fairly more evenly distributed among the PS, TR and AC spectral types, roughly in a 20--20--60 ratio, respectively. The Composite group is indeed the most populated of the narrow-line activity classes that contribute to the TR broad-band spectral type, whereas the members of the SF and Seyfert classes tend to avoid this region (they represent, respectively, only $\sim 2$ and $3$ per cent of the TR objects with a reliable BPT assignment). These last two spectral types are also, by far, the less well represented BPT classes in the PS ($\sim 0.5$ per cent each), a role that is played by the LINER class among the members of the AC ($\lesssim 4$ per cent).

\subsection{WHAN diagram}
\label{SS:WHAN}

According to \citet{Sta+08} one of the shortcomings of BPT diagrams is that they cannot discriminate between galaxies hosting a true weakly active nucleus (wAGN) -- to avoid confusion Seyferts are renamed as strong AGNs (sAGN) -- and `retired galaxies' (RG) \citep[see also, e.g.][]{Sin+13,Sta+15}. The latter are systems where star formation has stopped long ago, but when enough intergalactic gas is withheld, can show nebular emission lines arising from photoionisation by very hot low-mass evolved stars (HOLMES) during their short-lived post-AGB phase. The WHAN diagram proposed by \citet{CidF+10,CidF+11}, which depicts the \Ha\ EW versus the flux ratio $\Nii/\Ha$, is intended to solve this issue. This scheme contains the same basic demarcations SF/AGN and Seyfert/LINER of the BPT--NII diagram, conveniently transposed to the new coordinates, while relying on theoretical arguments about the ionising photons budget to partition the LINER region into wAGN and RG. For the last group, \citeauthor{CidF+11} proposed an upper $\mbox{EW}(\Ha)=3\,\angs$ boundary, regardless of the value of $\Nii/\Ha$. Objects mainly ionised by HOLMES, in turn, are further subdivided according to the presence or absence of emission lines in their spectra into ‘emission-line retired’ (ELR) galaxies, if they have an $\mbox{EW}(\Ha)\geq 0.5\,\angs$, and ‘line-less retired’ (LLR) galaxies when the \Ha\ line is weaker. However, for the present work we will consider RG as a single spectral group, since the distinction between ELR and LLR objects seems to obey more to limitations in the sensitivity of the detectors than to be physically motivated\footnote{It is quite a standard procedure in the literature to assume that emission lines with $\mbox{EW} < 0.5\,\angs$ are not convincingly detected \citep{Her+16}.}.

\begin{figure}
	\includegraphics[width=\columnwidth, viewport=1 1 360 530, clip=true]{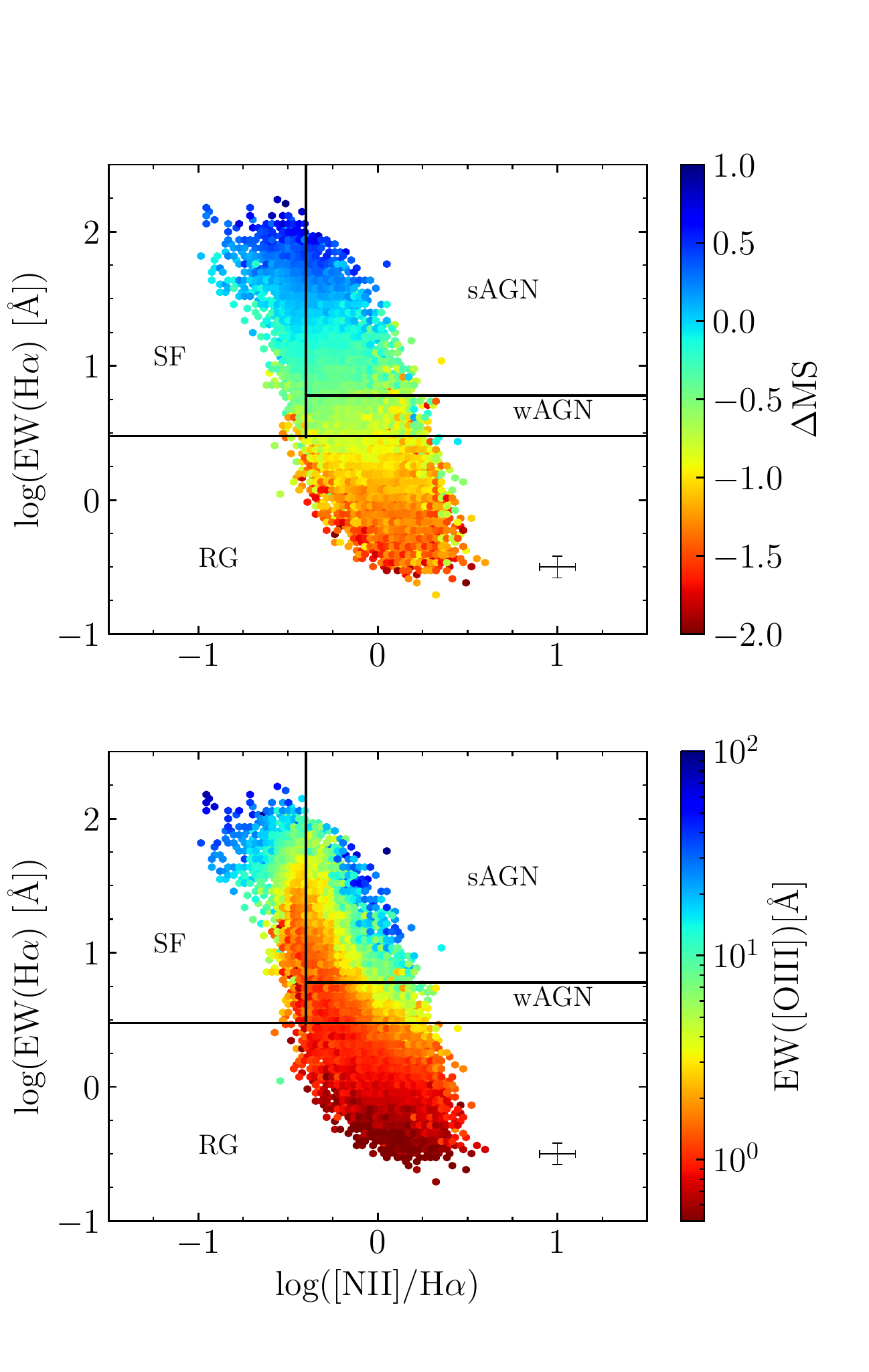}
	\vspace*{-1.5\baselineskip}
    \caption{Same as in Fig.~\ref{F:pc1-pc2_DMSO3}, but for the WHAN diagram. The horizontal black solid line at $\mbox{EW}(\Ha) = 6\,\angs$ represents the optimal transposition of the \citet{Kew+06} divider between sAGN and LINER derived in \citet{CidF+10}. A second horizontal division line at $\mbox{EW}(\Ha) = 3\,\angs$ separates the LINER zone into wAGN and RG (see text), while the vertical black solid line at $\log(\Nii/\Ha) = -0.4$ mimics the divider between SF galaxies (on the left) and all AGNs (on the right) of \citet{Sta+06}. The error bars in the bottom-right corners of the panels show the medians of the errors in both axes.}
    \label{F:whan_DMSO3}
\end{figure}

\begin{figure*}
	\includegraphics[width=\textwidth, viewport=110 -60 1300 310, clip=true]{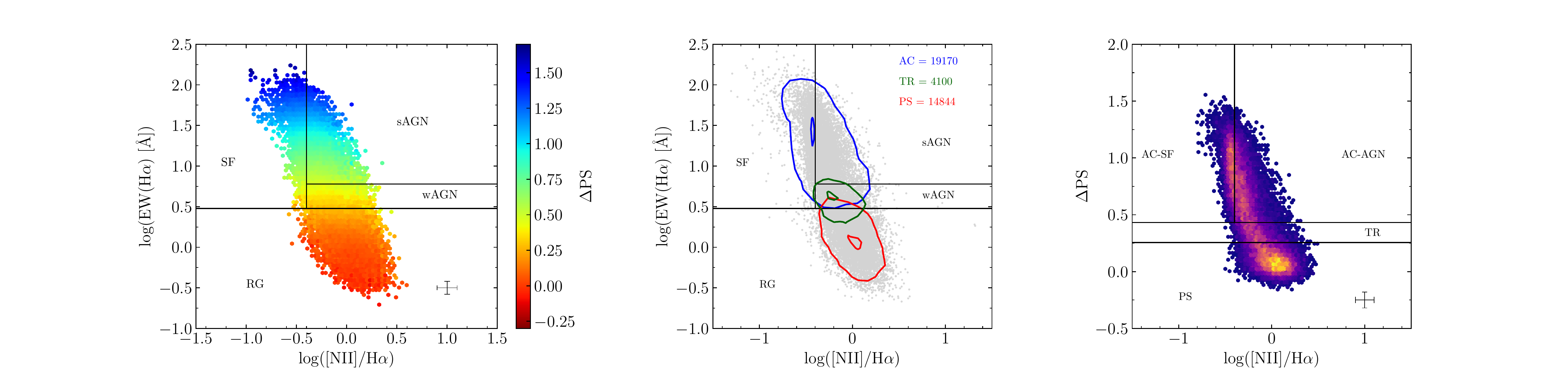}
	\vspace*{-3.0\baselineskip}
    \caption{\emph{Left:} Same as in Fig.~\ref{F:whan_DMSO3}, but using as third dimension the distance to the PS, \DPS, to show that it strongly correlates with EW(\Ha). \emph{Middle}: Same WHAN diagram with coloured contours corresponding to number densities of 10 and 90 per cent of the peak value for the PS (red), TR (green), and AC (blue) broad-band spectral types, with the actual numbers of galaxies in each category being listed in the upper-right corner of the panel. One in every two galaxies of the whole subset are plotted as light grey points in the background for reference. \emph{Right:} Heatmap of the DPSN diagram for our sample of nearby S0 galaxies. The vertical axis of the WHAN diagram is replaced in this new space by \DPS\ and the best linear approximations to the PS/TR and TR/AC boundaries are transposed into the horizontal divisors \DPS$\;=1.7$ and $2.8$, respectively. The SF/AGN vertical divider used by \citet{Sta+06} to separate these two spectral classes in terms of the ratio \Nii/\Ha\ has been preserved. These three plots highlight the close correspondence that exists between the emission-line classification introduced by \citet{CidF+11} for low-$z$ SDSS galaxies of all Hubble types and that defined in \citetalias{TSP20} from the PCA of the full optical spectra of the local S0 population.}
    \label{F:DPSN}
\end{figure*}

As done previously with the PC1--PC2 and BPT diagrams, we show in Fig.~\ref{F:whan_DMSO3} two representations of the WHAN diagnostic with colour gradients added to reveal any connection that may exist between the different spectral types and parameters \DMS\ (top panel) and EW(\Oiii) (bottom panel). We also include black straight lines to show the demarcations between the different spectral types discussed in the previous paragraph. By examining the upper panel of this figure, one can see that \DMS\ and EW(\Ha) are closely related, with the most actively star-forming galaxies dominating the upper area of the diagram and the most passive systems concentrating in the lower RG region. More specifically, the WHAN diagram reveals the presence of objects that possess a high SSFR in both the SF and sAGN (Seyferts according to the BPT naming) regions, a result which comes to show that sources powered by BH accretion can also exhibit important star formation activity. Similarly, nearly all of the most quiescent objects are located below the upper boundary of the RG region, $\mbox{EW}(\Ha) = 3\;\angs$, which concentrates $\sim 47$ per cent of all the S0s included in this graph. The comparison of the upper panels of Figs.~\ref{F:bpt_DMSO3} and \ref{F:whan_DMSO3} shows that these extreme values of \DMS\ encompass significantly wider ranges of the variable $\log(\Nii/\Ha)$ (x-axis) in the WHAN diagram than in the BPT one. Therefore, the differences that both layouts show in terms of their relationship with the parameter \DMS\ cannot be attributed to a possible reclassification of the activity of some galaxies. The most plausible explanation is the strong segregation of the SSFR imposed by the \Ha\ EW, which prevents galaxies with very different values of \DMS\ to intermingle, something that does happen, however, in the BPT--NII space. 

In addition, there is the constraint $\mbox{AoN} > 1.5$, which is applied to the four fluxes involved in the BPT--NII diagram to eliminate those sources with the most doubtful classifications, while for the WHAN diagram only the significance of the \Nii\ and \Ha\ emission-lines matters. This particular element allows us to classify another $15,\!319$ S0 galaxies using the WHAN diagnostic, the majority of whom have weak emission lines (only $\sim 1/3$ of the $17\!,893$ galaxies of the RG class have a BPT type assigned). Furthermore, had we taken a look at the probability density function (PDF) of the EW(\Ha) for the S0 galaxies designated SF on the BPT diagram, we would have seen that it follows a bimodal distribution, with the main mode centred on values characteristic of MS objects and a much smaller secondary peak located below $3\;\angs$ and encompassing values typical of RG-class objects, for which star formation is essentially dead and that should therefore be excluded from the SF class \citep[see e.g.][]{Fel17}.

On the other hand, the comparison of the bottom panels of these same figures shows that both the BPT and WHAN spaces feature distributions of the EW(\Oiii) variable that span an almost identical range with respect to the log(\Nii/\Ha) dimension. As regards the WHAN diagram, the highest values of this variable are attained in the upper-left and upper-right boundaries of the SF and sAGN regions, respectively, while the lowest values pile up mainly throughout the RG section. However, the correlation between the equivalent widths of \Oiii\ and \Ha\ is now very weak. In fact, we can observe in this latter activity classifier the existence of a wide wedge-shaped zone formed by objects with a very low \Oiii\ emission that protrudes above the RG region along the SF/AGN divider extending up to $\mbox{EW}(\Ha)\sim 50$. It is not unreasonable to expect that this \Oiii-devoid wedge, which covers a good part of the wAGN region and of the margins of the SF and sAGN regions closest to the \citeapos{Sta+06} divider, may comprise mainly galaxies whose activity is not necessarily related to the accretion into a central massive black hole. Interestingly, this would be consistent with the finding by \citet{Caz+18} that AGN is the dominant mechanism of ionization for LINERs when their flux ratios $\Nii/\Ha$ are $\gtrsim 0.7$. The implications of this result on the allocation of activity classes will be addressed in a separate work.

\subsection{DPSN diagram}
\label{SS:DPSN}

The WHAN diagram was introduced by \citet{CidF+10} as an alternative to BPT diagrams with a diagnostic power equal to or even better and with an improved economy of measurements that reduces the dependence on the sensitivity of the instrumentation and, therefore, also the number of emission-line galaxies excluded from line quality cuts and the effects of the resulting bias. Another two-dimensional activity classifier that preserves these good qualities can be obtained using for the $y$-axis the distance to the PS, \DPS, given by equation~(\ref{E:DPS}), and keeping the same horizontal dimension adopted in the WHAN and BPT--NII diagrams. This proposal is based on the fact that \DPS, which is essentially linked to the global shape of the optical continuum of the stellar population of the host galaxy, is also very strongly correlated with the EW(\Ha), as already shown in \citetalias{TSP20}. Since the PCA taxonomy can be reduced to a single component along the \DPS\ axis, it is expected that this new diagram can also provide a suitable topological separation for both the WHAN and PCA spectral classes.

These expectations are fully confirmed by Fig.~\ref{F:DPSN}. In the left-hand panel, we show a WHAN diagram in which the data have been coloured according to the average values of \DPS\ to corroborate that this parameter is strongly correlated with EW(\Ha), whereas in the middle panel we employ number density contours of different colours to trace the regions populated by the three PCA modes of lenticular galaxies in the WHAN subspace. While it could be anticipated from the layout of colours on the left-hand panel that most of the PS lenticulars would lie within the RG region and that the bulk of the members of the AC class would fall above the $\mbox{EW}(\Ha)=6\,\angs$ divider, i.e.\ in the SF+sAGN domains of the diagram, it is particularly telling that the distribution of TR members heavily sits within the boundaries of the wAGN section. Specifically, 95 per cent of the PS objects have $\mbox{EW}(\Ha)<3\,\angs$, 92 per cent of the ACs have $\mbox{EW}(\Ha)>6\,\angs$ and 73 per cent of the TRs fall between the 3 and $6\,\angs$ divisors. Note that the smaller fractions of the TR class can be explained by the uncertainties affecting the flux and EW estimates of narrow emission lines causing some data spillover through the dividing lines of the diagram, especially down and to the right of the plot, which is critical for a narrow area like that assigned to the wAGN class. This coincidence between divisor lines that rely on completely different spectral elements is all the more noteworthy considering that the PS/TR/AC classification emerges from purely mathematical arguments derived from the analysis of the first principal components of the optical spectra, while the thresholds applied to the EW of the \Ha\ line are empirical approximations to model-based and theoretically motivated physical frontiers. It can therefore be concluded that the taxonomic paradigm introduced in \citetalias{TSP20} from S0 galaxies strengthens the suitability of the EW(\Ha)-based divisors proposed by \citet{CidF+11} to distinguish wAGN from RG among the population of narrow emission-line galaxies. And vice versa, that the existence of dividers devised from the relative strength of the \Ha\ emission provides a physical basis for the broad-band PS/TR/AC spectral types inferred in \citetalias{TSP20}.

Based on the above considerations, we introduce in the right-hand panel of Fig.~\ref{F:DPSN} a new activity diagnostic diagram, which we will call DPSN, using our sample of nearby S0 galaxies as a test bed. In this scheme, the PS/TR and TR/AC demarcations included in Figs.~\ref{F:pc1-pc2_dens} and \ref{F:pc1-pc2_DMSO3} are converted, respectively, into the horizontal divisor lines \DPS$\,=1.7$ and $2.8$, while the optimal SF/AGN vertical separation of \citet{Sta+06} is kept. The DPSN subspace, which essentially preserves the shape and relative locations of the SF, sAGN, wAGN, and RG areas from the WHAN diagnostic, maintains a similar discrimination ability in the vertical direction, but now makes it more resistant to measurement errors, since the location of the sources along this axis is set by a parameter that relies upon the flux contained in their entire optical spectra rather than on a single line. In addition, once the base of eigenvectors has been determined, something that must be done only once for a given spectroscopic galaxy sample, the calculation of the PC of the individual spectra, and hence of \DPS, is relatively straightforward, making unnecessary using stellar synthesis analysis to subtract the continuum flux. Other advantages of using this latter parameter instead of EW(\Ha) to measure activity are that the distribution of its values is barely affected, in a statistical sense, by both internal extinction and fixed aperture effects (see \citetalias{TSP20}), and, although it does not have a practical impact on the census of galaxies qualifying for classification -- the horizontal dimension continues to provide the strictest quality cut --, that it is possible to infer the distance to the PS for any object endowed with a good-quality optical spectrum, regardless of the host's level of activity. The latter is particularly useful for samples, such as the one we are studying here, in which galaxies with absorption spectra abound. 

In the next two sections, we discuss the properties of the various spectral classes of local S0s in regions of the electromagnetic spectrum, other than the visible, using data drawn from some of the most important mid-IR, radio and X-ray wide-area surveys. We want thereby to increase our understanding of the different subpopulations of lenticular galaxies by examining them from different angles and, ultimately, unravel their origin.

\section{Infrared Detections of Present-day S0 galaxies}
\label{S:midIR}

An important issue that affects any diagnostic diagram for galaxies that uses fluxes and/or EWs is that these parameters are aperture-dependent. This has led several authors \citep[e.g.][]{Her+16,Zew+20} to consider alternative classification paradigms that rely partially or totally on photometric data. This is justified on the basis that differences in the astrophysical sources driving galaxy evolution translate to differences in the SED and hence in the colours of the hosts. 

It soon became clear that the mid-IR waveband was particularly suitable to search for such differences. One of the most powerful diagnostics currently available in this electromagnetic window are colours derived using photometry from the \textit{Wide-Field Infrared Sky Explorer} \citep[\emph{WISE};][]{Wri+10} all-sky survey. This observatory has mapped the entire sky in four mid-IR broad bands \wi, \wii, \wiii, and \wiv\ centred at 3.4, 4.6, 12, and $22~\mic$, respectively. Even though the study of AGNs was not among the main goals of \emph{WISE}, this survey is really appropriate for that aim since its mid-IR detectors roughly cover one of the wavelength ranges in which AGNs can be clearly distinguished from SF galaxies \citep[see e.g.\ fig.~1 in][]{HA18}. In particular, \emph{WISE} observations are useful in discovering AGNs whose emission in the UV-optical waveband has been obscured by the dust and gas within the torus surrounding the accretion disc of the SMBH and/or by dust in the galaxy \citetext{e.g.\ \citealt{Ste+12,Mat+13,Ass+13}; see also the review on obscured AGNs by \citealt{HA18}, as well as next section}. However, it is important to remember that mid-IR observations are only sensitive to relatively luminous AGNs in which nuclear emission is strong enough to dominate over emission from other physical processes in the host galaxy, such as star formation, which in extreme cases (e.g.\ starbursts) can dilute or overwhelm the AGN output \citep{GHJ14,Rad+21}. Another advantage of using colours to classify galaxies is that they can be inferred for any type of object, whether it has an emission spectrum or not.

Next, we focus on the S0s in our data set that are also detected in the AllWISE Source Catalog, which lists the most sensitive and accurate data released by the \emph{WISE} survey, including, among other products, astrometry and photometry for over 747 million objects.

\begin{table}
\centering
\caption{Joint distribution of the PCA and BPT spectral classes for S0 galaxies with $z\leq 0.1$ detected in the AllWISE catalog.}
\label{T:BPT-PCA_midIR}
\begin{tabular}{cccc} 
\hline
		BPT class &  & PCA type& \\
		 & Active Cloud & Transition Region & Passive Sequence \\
		\hline
Star-forming &1744&19&8\\
Seyfert &262&23&10\\
LINER &211&214&1024\\
Composite &1153&495&691\\ 
Unclass &495&375&5202\\
\hline
\end{tabular}
\end{table}

\subsection{Crossmatch to AllWISE catalog}
\label{SS:AllWISE}

The AllWISE counterparts of our S0 galaxies have been extracted by selecting the closest pair within a search radius of 2 arcsec, a sensibly shorter distance than the $6.1$ arcsec angular resolution of the \wi\ band. Despite the strictness of the rule applied, we do not expect to miss a significant number of pairs, as a previous test demonstrated that the distribution of angular separations for matched sources is strongly skewed towards values $<1$ arcsec. Besides, we have excluded AllWISE detections that are consistent with point sources, i.e.\ having an extended source flag parameter \texttt{ext\_flg}$\,=0$, to eliminate potential matches with stellar objects. In an effort to achieve a mid-IR sample of the highest quality, we have also demanded a minimum SNR of 5 in the \wi\ and \wii\ bands, and of 3 for the less sensitive \wiii\  magnitude\footnote{We do not make use in this work of the much shallower and noisier \wiv\ band data.}. Finally, only AllWISE counterparts with a good quality photometry, i.e.\ having \texttt{cc\_flag}$\,=0$ in the first three bands, have been considered. 

After applying all the aforementioned constraints, we obtain a mid-IR sample with a total of 11,926 present-day S0 galaxies. While this figure represents an overall reduction of $\sim 5$:1 from the size of the baseline optical database, the relative percentage abundances of the different activity groups barely register, in general, significant changes (see Table~\ref{T:BPT-PCA_midIR}). Thus, in the \emph{WISE} energy band the SF and Seyfert continue en masse being part of the AC region, maintaining percentages of around 99 and 90 per cent, respectively, with the LINER and Unclass type of lenticulars being mostly members of the PS, and the Composite the most evenly distributed BPT class within the PCA broad-band categories, while the global fraction of objects without a BPT classification drops slightly to 51 per cent.

\subsection{Mid-IR colour--colour diagrams}
\label{ccdiagrams}
The WHAN diagram (section~\ref{SS:WHAN}), just like the proposed DPSN alternative (section~\ref{SS:DPSN}), enable rather exhaustive descriptions of the activity exhibited by galaxies because they deal with a pair of the strongest emission lines in the optical window and have their vertical axes based on measurements that are scarcely affected by the quality of the data. However, such subspaces are not entirely free of sensitivity issues, specially regarding their horizontal dimension. 

The WHAW diagram \citet{Her+16} constitutes one attempt to improve over the WHAN classification by replacing the $x$-axis coordinate with the \wxxiii\ colour. Unlike emission line ratios, colours are global parameters applicable to galaxies of any kind and, in particular, \wxxiii\ is expected to do a very good job at separating objects with a high SSFR from RG \citep[e.g.][]{Clu+14,Ore+17}. However, such a change in the horizontal axis is not accompanied by a clear differentiation between SF and Seyfert galaxies, at least not as much as that provided by the BPT- and WHAN-like spaces. This has restricted the applicability of this type of hybrid diagnostic diagram that combines optical spectroscopy with mid-IR photometry, and might explain why it has been scarcely used in the literature.

A different approach for galaxy classification is provided by fully photometric schematics that, thanks to their insensitivity to aperture effects, enable the categorisation of all types of sources, including galaxies totally devoid of emission lines. \emph{WISE} colour--colour plots keep the \wxxiii\ colour of the horizontal axis used by the WHAW diagrams, but replace the vertical coordinate by the \wxii\ colour to achieve a sound topological separation of sources that show certain levels of activity, such as AGNs and ultra-luminous infrared galaxies (ULIRGs)\footnote{These objects are usually absent in data sets that, like the one that concerns us here, are based on optically selected galaxies.}.

\begin{figure}
	\includegraphics[width=\columnwidth, viewport=1 10 440 650, clip=true]{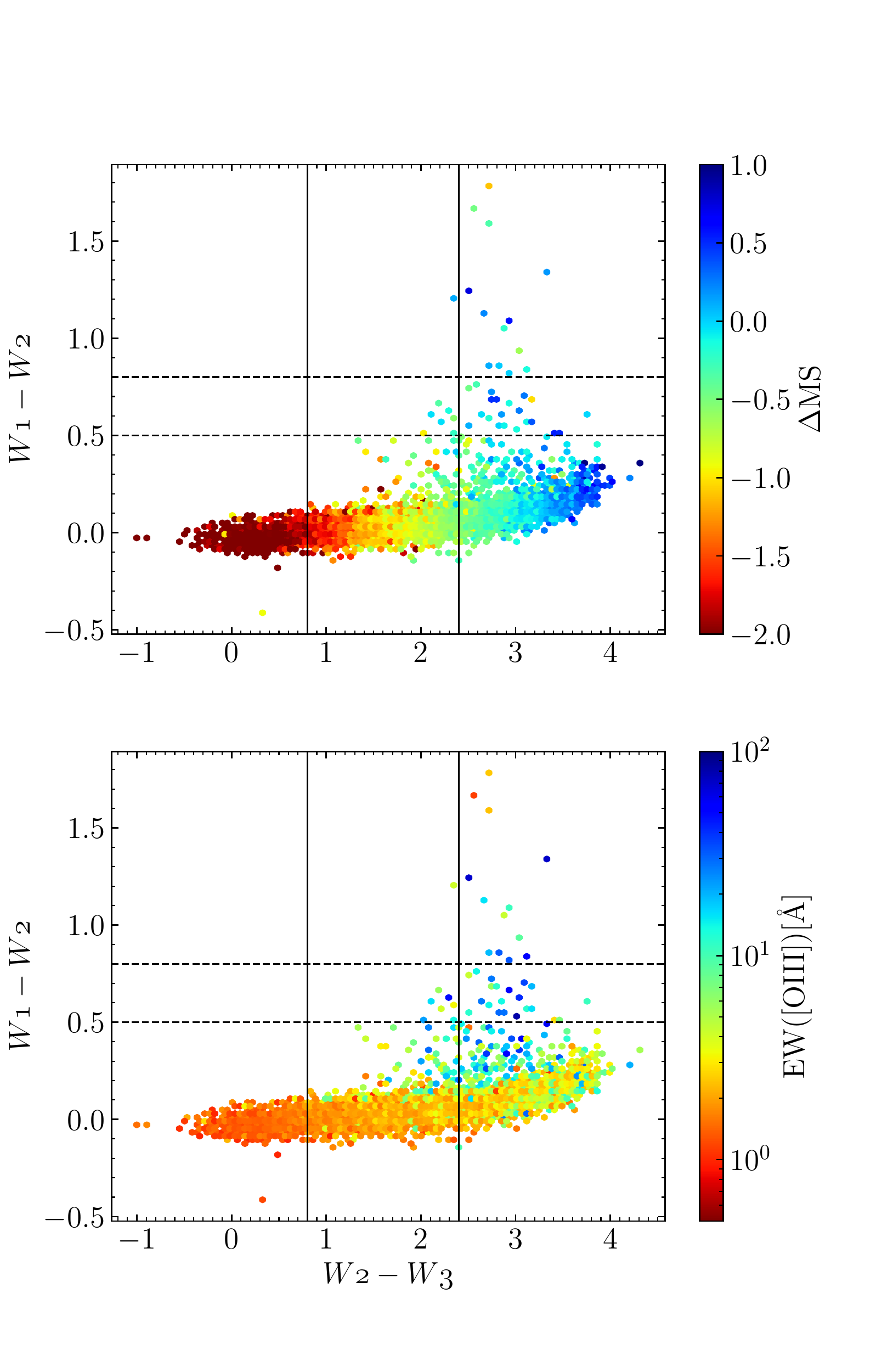}
	\vspace*{-2.0\baselineskip}
    \caption{Same as in Fig.~\ref{F:pc1-pc2_DMSO3} but for the \wxii\ vs.\ \wxxiii\ colour--colour diagram of the members of our S0 sample with a counterpart in the AllWISE catalog. The two vertical black solid lines demarcate the edges of \emph{WISE}'s IRTZ defined to be \wxxiii$\,\approx\,0.8$--$2.4$ by \citet{Ala+14}, while the \wxii$\,>0.5$ and the stricter \wxii$\,>0.8$ colour cuts adopted, respectively, by \citet{Ass+13} and \citet{Ste+12} to select AGN sources, are shown as horizontal black dashed lines. Despite the important difference in sample sizes, we have kept the same ranges of the colour bars used in previous diagrams for ease of comparison.}
    \label{F:cc_DMSO3}
\end{figure}

\begin{figure*}
	\includegraphics[width=\textwidth, viewport=35 -25 570 380, clip=true]{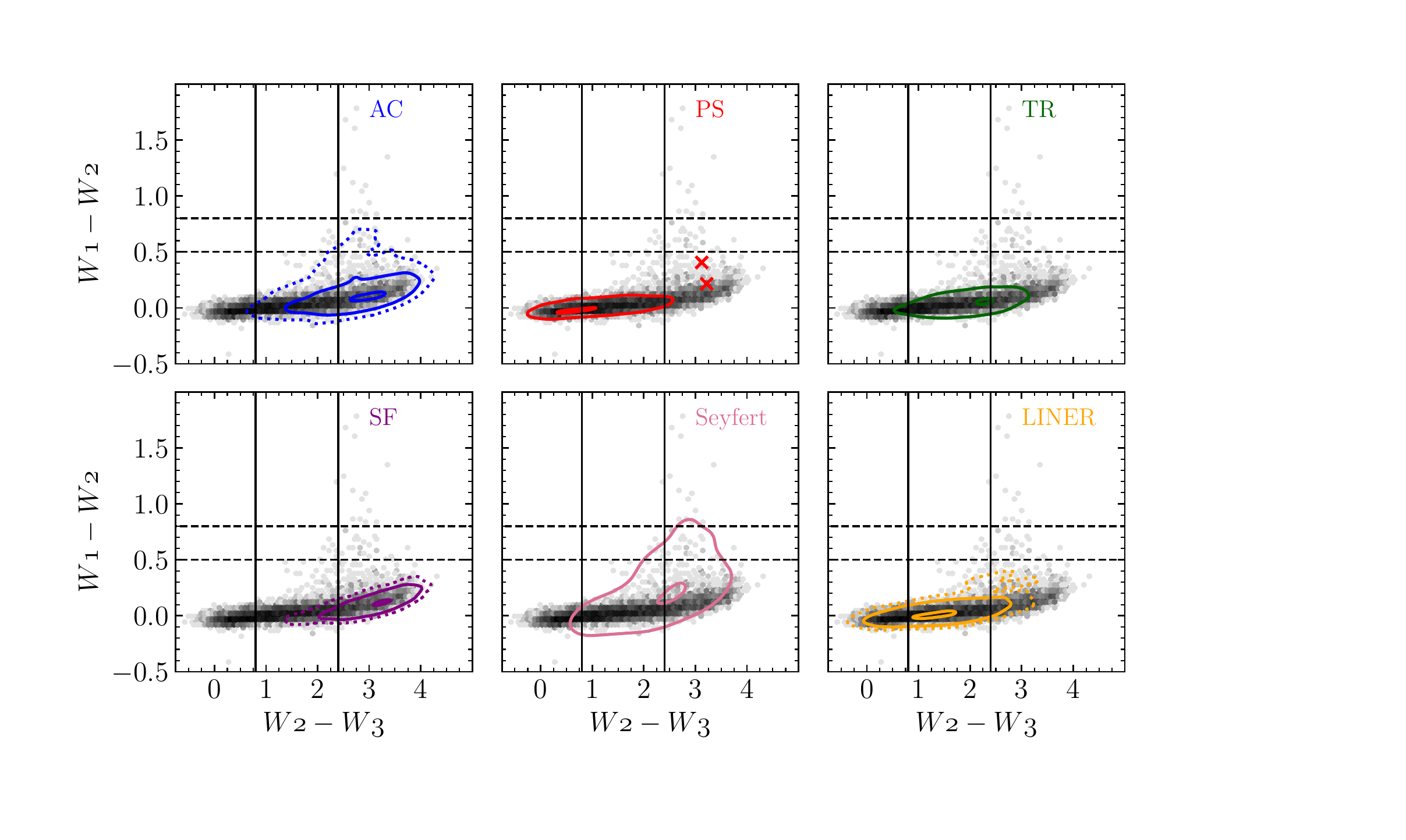}
	\vspace*{-7.5\baselineskip}
    \caption{Mid-IR colour--colour diagrams for the main broad-band and narrow-band optical spectral classes of the nearby S0s included in AllWISE. From left to right and top to bottom: AC lenticulars (blue), PS (red), TR (green), SF (purple), Seyfert (pink), and LINER (orange). Contours in each panel are drawn at number densities 10 and 90 per cent of the peak value. Extra contours (dotted lines) are added to some panels at number densities one per cent (AC) and 3 per cent (SF and LINER) of the peak values to show better the total breadth of these distributions. Two PS galaxies with abnormally red \wxxiii\ colours discussed in the text are marked with red crosses. In all panels, a heatmap of the full AllWISE S0 sample is shown in the background using a logarithmic greyscale. The vertical black solid lines are the limits of the IRTZ, while the horizontal black dashed lines show the lower \wxii\ colour cuts adopted by \citet{Ass+13} (bottom) and \citet{Ste+12} (top) to select AGN sources.}
    \label{F:cc_vs_classes}
\end{figure*}

Fig.~\ref{F:cc_DMSO3} displays the \wxii\ vs.\ \wxxiii\ colour space for the subset of nearby S0s detected in AllWISE. As in previous figures, we have coloured the data points according to the value of parameters \DMS\ (top panel) and EW(\Oiii) (bottom panel). The two plots reveal (see also Fig.~\ref{F:cc_vs_classes}) that most measurements are distributed over a narrow and slightly bent upwards strip that runs roughly parallel to the horizontal axis. The upper right part of this band connects with a sparsely populated region -- outlined by hexagonal cells that typically contain a single data point -- that extends across the reddest sectors of the plot and hosts a sizeable fraction of lenticulars with important levels of activity (see below). The much more crowded banana-shaped region encompasses roughly the same range of IR colours displayed by data sets of nearby galaxies spanning the entire Hubble Sequence. However, and not unexpectedly, the distribution shown by the S0s is predominantly consistent with those sectors of the diagram traditionally associated with early-type systems, lacking, for example, the deep gap in the range $\mbox{\wxxiii}\approx\,0.8$–$2.4$ mag, deemed as the infrared transition zone (IRTZ), that appears in samples that contain ETGs and LTGs  \citetext{see, e.g., fig.\ 1c of \citealt{Ala+14} for a detailed view of this area}. Regarding the values of \DMS, and hence of the SSFR, we find that they are strongly correlated with the \wxxiii\ colour, so that all but five of our S0 systems with enhanced star formation, i.e.\ having \DMS$\,\gtrsim 0$, are housed in sectors redwards of the upper boundary of the IRTZ. Something similar happens to the values of EW(\Oiii), which also increase with the mid-IR colours of the \emph{WISE} space, although in this case with a substantially larger dispersion that makes it pretty easy to find objects with strong \Oiii\ emission ($\mbox{EW}(\Oiii)\gtrsim 10\,\angs$) within the IRTZ and that considerably intermingles the values of this parameter in said space. In contrast, those lenticulars that are passive in all respects cluster bluewards of the IRTZ -- notice that the third dimension of the panels assigns them the reddest cells -- and exhibit \wxxiii\ colours near zero \citep{Jar+11}.

Among the S0s with enhanced \wxii\ colour, we find 44 sources that verify the condition $\mbox{\wxii} > 0.5$, the criterion adopted by \citet{Ass+13} to obtain highly reliable AGN samples \citetext{see also \citealt{Ste+12} for a somewhat stricter threshold}, whereas very few of them, three to be exact, show values of $\mbox{\wxii}\gtrsim 1.6$ characteristic of obscured AGN. Forty of the galaxies above the \citeapos{Ass+13} colour cut (91 per cent of the total) belong to the AC class, 27 of them tagged as Seyfert (57 per cent) and only one as SF. It thus appears that the \citeapos{Ass+13} limit provides, at least for present-day S0 galaxies, not only a good criterion to obtain reliable samples of systems powered by nuclear SMBH, but also an upper boundary to the \wxii\ colour that can be achieved by these objects when the principal source of activity comes from star formation.

Fig.~\ref{F:cc_vs_classes} depicts the location of the spectral classes defined through the PCA and BPT diagnostics within the \emph{WISE} diagram\footnote{Although this and the next comparisons with the PCA types have been restricted for space reasons to the main BPT classes, they are also representative of the corresponding WHAN classes.}. With regard to the PCA types, we find that AC lenticulars are primarily located to the right of the upper boundary of the IRTZ, in the lower right sector of the diagram, although we can also see an extended low-density belt going further into the IRTZ and upwards to the AGN region. The reddest divider of the IRTZ also acts as an upper limit for nearly all PS objects, which are essentially found to the left of it. In turn, a sizeable number of the TR members lie within both edges of the IRTZ, with the peak of their distribution at the red boundary of this region, reinforcing the idea that this spectral class encompasses objects evolving towards a quiescent state. In terms of the BPT classes, we find that S0--Seyferts occupy a region not unlike that of the AC members, albeit somewhat more extended along the vertical direction, covering the entire IRTZ and reaching redder \wxii\ colours. While it is true that some of these objects also exhibit high values of \DMS, the fact that the Seyfert spectral class shows a far larger spread in the \wxii\ vs.\ \wxxiii\ colour space than the S0--SF -- the largest of all spectral groups indeed -- suggests that in a number of cases its members owe their mid-IR colours mainly to the presence of nuclear engines in the hosts \citep[see also, e.g.][]{Ala+14}. Another BPT class of S0 galaxies that tends to scatter in the \emph{WISE} colour space towards the more active areas of it (although less so than Seyferts and ACs), is the LINER, probably for the same reason that we have just commented. This class has also a good number of objects in common with the TR one (see Table~\ref{T:BPT-PCA_midIR}), both spectral groups being the main dwellers of the IRTZ, but with the distribution of the former more centred in this area, which seems to indicate that the LINER class encompasses objects with somewhat lower activity levels. For its part, the few lenticular galaxies that are actively forming stars are preferentially found redwards of the upper edge of the IRTZ, although there is also a sparse tail that penetrates into this region up to $\mbox{\wxxiii}\sim 1.5$ in accordance with the findings of \citet{Clu+14} from the resolved sources of the Galaxy And Mass Assembly (GAMA) survey. This suggests that, in spite of their SF status, some members of this class are already beginning to see their SFR diminished. Generally speaking, we can conclude that the loci occupied by the different spectral classes of present-day lenticulars in the \wxii\ vs.\ \wxxiii\ diagnostic diagram show a good agreement with those that can be inferred for the entire local population of galaxies \citep[compare our Fig.~\ref{F:cc_vs_classes} with, e.g.\ fig.\ 7 in][]{Zew+20}, the most notable difference being that the \emph{WISE} colours of our sample are bluer on average. This is unsurprising given the placement of S0 galaxies within the Hubble Sequence.

We shall end this section by discussing two lenticular galaxies that have come to our attention because in spite of their PS classification are found well past the upper limit of the IRTZ (red crosses in the PS panel of Fig.~\ref{F:cc_vs_classes}). Close inspection of the spectrophotometric data available for these two objects (\texttt{specObjID} = 1083174028455208960 and 395248620099627008) in the \emph{SciServer} of the SDSS reveals that they have very red stellar continuum spectra, which explains their PCA classification as passive systems, while showing at the same time relatively prominent nebular emission lines. The red SDSS colours, which together with the high intrinsic brightness justify their selection as Luminous Red Galaxies, appear to be the result of a combination of closeness ($z \sim 0.03$), high inclination ($i \sim 60\,\mbox{deg}$), and presence of relatively significant amounts of gas and dust. This last factor points to the existence of a certain level of star formation in these objects capable of generating emission lines, but whose main booster is most likely nuclear activity, since both S0s are cataloged in the literature as radio sources with post-starburst stellar populations, in agreement with their BPT rating as Seyfert and Composite, respectively. In any case, these seemingly contradictory spectral and photometric diagnoses can be perfectly understood, while their complementarity is revealed. More specific statements about the evolutionary status of these and other galaxies included in our mid-IR S0 sample would require follow-up studies of their baryonic content that are beyond the scope of this paper.

\section{Radio and X-ray Detections of Present-day S0 galaxies}
\label{S:radio_X-ray}

Radio wavelengths are widely used to characterise nuclear emission. In fact, AGNs were first discovered by their luminosity in the radio waveband. The emission in this regime from high-luminosity radio galaxies is dominated by non-thermal synchrotron radiation arising from processes related to the accretion disc around the central SMBH and/or powerful relativistic jets. These radio-loud sources comprise only a small fraction of the general AGN population, which is predominantly radio quiet \citep[e.g.][and references therein]{ZSM08,Kel+16}. Synchrotron emission by relativistic electrons, accelerated in this case in supernova shocks, is also one of the two main components of the radio continuum emission related to the evolution of massive stars. The other is thermal free–free radiation from ionised gas in the \Hii\ regions surrounding them \citep{Con92}. Besides, the optical depth for radio emission is very low and so the 20-cm (1.4 GHz) radio continuum can be used to observe both the star formation rate in dusty galaxies and the most obscured AGNs that can escape detection in the optical, mid-IR and X-ray windows. To differentiate between non-thermal activity and that related to star formation one can require $L_{\mathrm{1.4\;GHz}}\gtrsim 10^{23}\;\mbox{W}\,\mbox{Hz}^{-1}$, which is roughly the luminosity at which radio source counts switch from being largely dominated by star-forming galaxies to being dominated by powerful AGNs \citetext{see e.g.\ the radio luminosity function in fig.\ 11 of \citealt{Bes+05}}. Below this limit, AGNs cannot be reliably distinguished from SF on the basis of radio luminosity alone. However, by assuming that for local $L\gtrsim L^\ast$ galaxies non-thermal radio emission directly tracks their SFR, we can use the latter to calibrate the radio power at $1.4$~GHz according to the empirical relation by \citet{Bel03}\footnote{This calibration only considers the contribution of synchrotron emission from relativistic electrons associated with SN remnants which, according to the standard model of SF luminous galaxies, accounts for $\sim 90$ per cent of the total radio continuum flux at 1.4 GHz \citep{Con92}.}
\begin{equation}\label{sfr-radio}
L^{\mathrm{SFR}}_{\mathrm{1.4\;GHz}} = 1.8\cdot 10^{21}\left[\frac{\mbox{SFR}}{\mbox{M}_\odot\;\mbox{yr}^{-1}}\right]\;\mbox{W\,Hz}^{-1}\;,
\end{equation}
the scatter being a factor of around $\pm\,0.26$ dex on a galaxy-by-galaxy basis \citep[see also][]{YRC01}.

AGN activity can also be traced through X-ray emission and is believed to be produced by Comptonisation of optical/UV photons of the accretion disc by the corona of hot electrons that envelops it \citep[e.g.][]{HM91}. These observations, however, can undergo some absorption, especially in thicker Compton sources with $N_{\mathrm H} > 10^{24}\;\mbox{cm}^{-2}$ \citep[e.g.][]{DiaS+09,LaM+09}. As in the radio domain, stellar processes related to diffuse gas and X-ray binaries in galaxies can also be a source of X-ray emission. A sensible criterion to disentangle the type of source that dominates the X-ray counts, AGN or stellar, is provided by the \citeapos{RCS03} relation between SFR and hard (2--10 keV) X-ray luminosity calibrated for local samples \citep[c.f.][]{TB10}:
\begin{equation}\label{sfr-Xray}
L^{\mathrm{SFR}}_{\mathrm{2-10\;keV}} = 5\cdot 10^{39}\left[\frac{\mbox{SFR}}{\mbox{M}_\odot\;\mbox{yr}^{-1}}\right]\;\mbox{erg\,s}^{-1}\;,
\end{equation}
which implies that a SFR of at least $200\;\mbox{M}_\odot\,\mbox{yr}^{-1}$ is required to produce an X-ray luminosity greater than $10^{42}\,\mbox{erg\,s}^{-1}$ \citetext{\citealt{PR07} and \citealt{Vat+12} find similar relations between the X-ray hard luminosity and the SFR within a factor of $\sim 1.5$ of equation~(\ref{sfr-Xray})}. These huge SFRs are typical of ULIRGs, and we expect none in our S0 sample (see section~\ref{ccdiagrams}). Therefore, any source in our data set more luminous than this limit would likely be an AGN on energetic grounds. Again, weak AGNs may not be correctly identified by just applying this constraint.

\subsection{Crossmatch to FIRST and 4XMM-DR11/CSC 2.0 catalogs}
\label{SS:radio_X-ray_match}

The VLA Faint Images of the Radio Sky at Twenty centimetres (FIRST) Survey \citep{BWH95} covers 10,575 deg$^2$ of the radio sky over the north and south galactic caps by using the Very Large Array in its B configuration. This results in a high image resolution of 5--6 arcsec and a source position accuracy of $< 1$ arcsec. The FIRST footprint was designed to overlap with the sky area surveyed by the SDSS, so $\sim 30$ per cent of the sources have optical counterparts in the Sloan data. The sensitivity of the observations is also excellent, with a limiting flux density of about one mJy for point sources. We select from the FIRST catalog those objects with an S0 counterpart in our baseline data set within a fiducial radius of 8 arcsec, eliminating stellar sources (\texttt{SDSS\_class==’s’}) and those that are likely spurious (\texttt{P(S)\;>\;0.1}). We also demand the ratio between the radial separation $r$ and the $R_{\mathrm 50}$ Petrosian radius of our S0 galaxies to be lower than 1 as a way of increasing the probability that we are selecting the same source in both data sets. With this procedure we obtain a total of $1586$ matches, including $1176$ unique and 410 multiple, for which we have chosen the narrowest pair.

For the crossmatch with X-ray data, we use two catalogs in an attempt to remedy somewhat the significantly poorer statistics offered by the shallow observations in this waveband. On the one hand, there is 4XMM-DR11 \citep{Web+20}, the year 2020 version of the fourth generation of the catalog of Serendipitous X-ray sources from the \emph{XMM--Newton} telescope \citep{Jan+01}. Although not an all-sky survey, this is actually the largest X-ray catalog in existence, comprising 895,415 individual detections of 602,543 unique X-ray sources in the energy range from $0.2$ to 12~keV distributed over 1239 deg$^2$ of the sky and with a median positional accuracy for the point source detections of $< 1.57$ arcsec. We have matched our S0 data set to sources in this catalog located at an angular separation of less than 10 arcsec and used the information listed in the detection flags to eliminate possible spurious and/or contaminated sources, producing a first subset of 298 S0 galaxies with measured X-ray fluxes.

On the other hand, we have also crossmatched our S0 with the final second major release of the \emph{Chandra} X-ray observatory \citep{Sch04}. The Chandra Source Catalog \citetext{CSC $2.0$; \citealt{Eva+20}} lists measured properties of 317,167 unique point and extended compact sources distributed across the entire sky (but not covering it) and observed with a position error of $0.71$ arcsec (95 per cent confidence) for $\sim 90$ per cent of them. \emph{Chandra}'s Advanced CCD Imaging Spectrometer provides fluxes in five science energy bands from $0.5$ to $7.0$~keV, including the broad-band master fluxes that cover the entire range of photon energies. To carry out the pairing with \emph{Chandra}'s data, we have taken advantage of the CSC $2.0$-SDSS table of unambiguous matches that lists, among other parameters, the source type, the object identifier and coordinates from SDSS-DR15, the angular distance $r$ between the matched sources, and the match probability. We accept as bona fide matches those with an equal \texttt{ObjID} as long as they are at a radial separation $r\leq 5$ arcsec and have a minimal match probability of $0.9$. This has produced a second subset of 123 matches, less than half of those obtained with \emph{XMM--Newton} catalog. After removing the 29 sources common to both subsets, we obtain a net total sample of 392 nearby X-ray emitting S0 galaxies, a quantity that, while not small, is well below the sizes reached by the samples inferred from the other bandwidths.

\subsection{Characterization of radio and X-ray detections}
\label{SS:radio_X-ray_mainfeatures}

\begin{figure*}
	\includegraphics[width=\textwidth, viewport=65 0 870 540, clip=true]{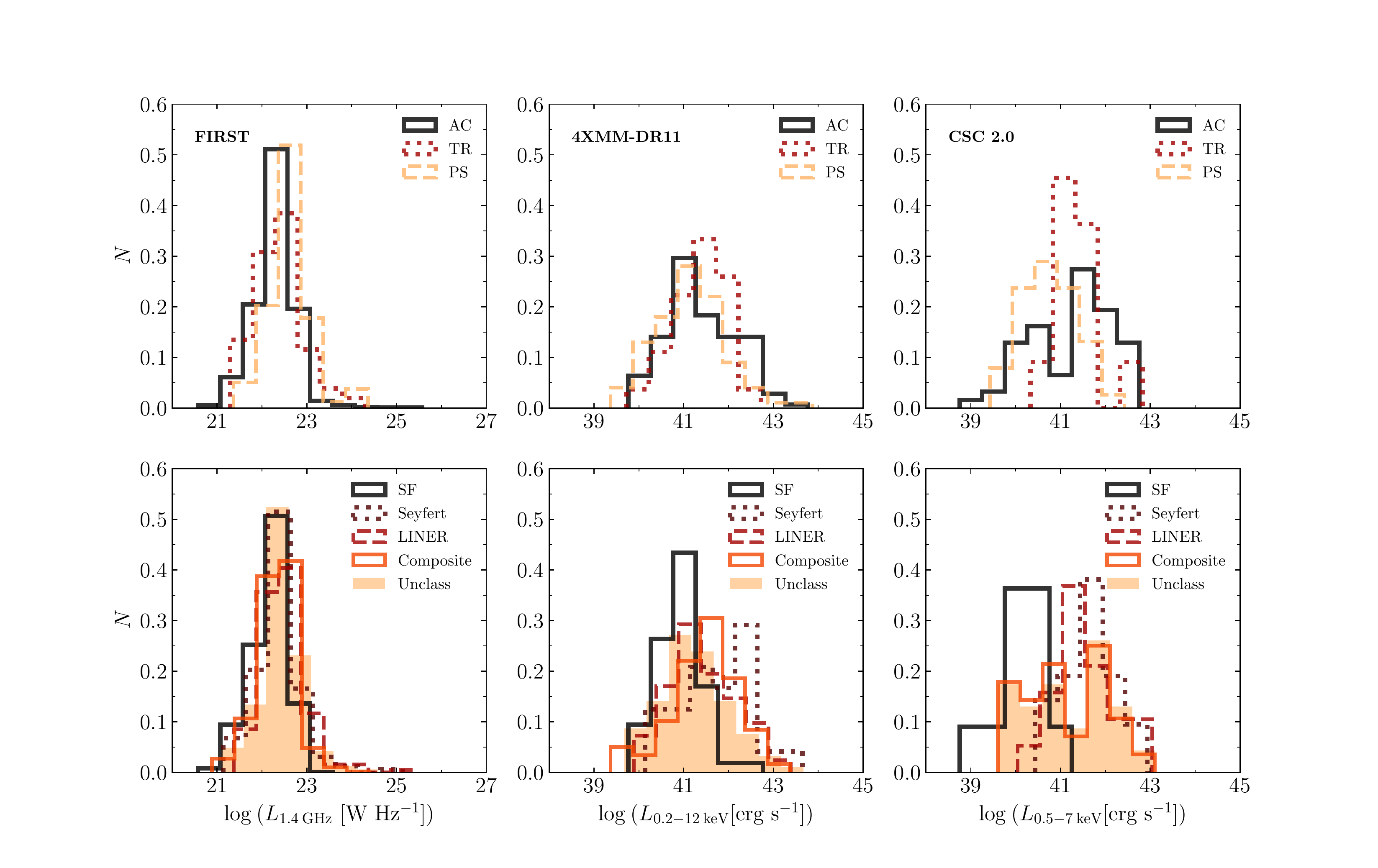}
	\vspace*{-2.0\baselineskip}
    \caption{Radio and X-ray luminosity frequency distributions for the PCA (\emph{top}) and BPT classes (\emph{bottom}) of present-day S0 galaxies. The 1.4 GHz radio luminosities are inferred from the monochromatic flux integrated over the source listed in the FIRST catalog (\emph{left}), while the X-ray luminosities have been calculated from the broad-band fluxes included in the 4XMM-DR11 (\emph{middle}) and CSC $2.0$ (\emph{right}) catalogs. Both radio and X-ray luminosities are uncorrected for absorption; the low $z$ of the sources also makes it unnecessary to apply rest-frame and other cosmological corrections. The height of the histograms represents the relative fraction of the members of a given spectral class in the different bins defined for each instrument in the $x$-axis.}
    \label{F:Radio_X-ray_lum}
\end{figure*}

We show in Table~\ref{T:BPT-PCA_radio} the bivariate distribution of FIRST S0 detections in terms of the PCA and BPT spectral classes, while in Table~\ref{T:BPT-PCA_Xray} we do the same for those S0s that are X-ray emitters. Not unexpectedly, most galaxies that appear in both tables are preferentially grouped into spectral classes associated with some type of activity. In the radio domain, objects of the SF class still dominate the statistics among the BPT spectral groups, with the most noticeable change with respect to the data listed in Table~\ref{T:BPT-PCA_vis} corresponding to the LINERs. Within this group, sources simultaneously tagged AC are clearly more abundant (by more than a 5:1 ratio) than those belonging to the PS class, thus reversing the trend observed at visible wavelengths. This major shift is largely responsible for strongly skewing the membership of the PCA spectral groups towards the AC class, which now accounts for more than 90 per cent of all detections. Although these outcomes point to a greater prevalence, in relative terms, of galaxies with nuclear activity, we would not be talking in any case about radio-loud AGN\footnote{The generally low power of the sources in our sample warrants that the sizes of most of them are likely close to or below the resolution limit and therefore that their measured radio fluxes are accurate.}, as can be deduced by observing that the distributions of the $1.4$~GHz continuum radio luminosity of the various spectral classes reflected in the left-hand panels of Fig.~\ref{F:Radio_X-ray_lum} softly cap at $\sim 10^{23}\;\mbox{W}\,\mbox{Hz}^{-1}$. Furthermore, the scant differences shown by these distributions could be considered to support claims that radio emission in radio-quiet AGNs is mainly due to star formation, and therefore that the radio power is also a good tracer of the SFR in galaxies of this spectral class \citep[e.g.][]{Bon+15}. This automatically would imply that the vast majority of local S0--Seyferts are in hosts that are also actively forming stars; we return to this issue below. 

\begin{table}
\centering
\caption{Joint distribution of the PCA and BPT spectral classes for S0 galaxies with $z\leq 0.1$ detected in the FIRST catalog.}
\label{T:BPT-PCA_radio}
\begin{tabular}{cccc} 
\hline
		BPT class &  & PCA type& \\
		 & Active Cloud & Transition Region & Passive Sequence \\
		\hline
Star-forming &594&0&0\\
Seyfert &160&2&1\\
LINER &139&25&24\\
Composite &458&11&8\\ 
Unclass &104&14&46\\
\hline
\end{tabular}
\end{table}

\begin{table}
\centering
\caption{Joint distribution of the PCA and BPT spectral classes for S0 galaxies with $z\leq 0.1$ detected in the 4XMM-DR11 and/or CSC $2.0$ catalogs.}
\label{T:BPT-PCA_Xray}
\begin{tabular}{cccc} 
\hline
		BPT class &  & PCA type& \\
		 & Active Cloud & Transition Region & Passive Sequence \\
		\hline
Star-forming &66&0&0\\
Seyfert &44&1&0\\
LINER &31&12&29\\
Composite &65&14&11\\ 
Unclass&13&12&104\\
\hline
\end{tabular}
\end{table}

The scarcity of sources with powerful nuclear emissions in the subset of X-ray detections is corroborated by the finding of a relatively small fraction ($16.6$ per cent) of galaxies whose broad-band luminosity exceeds $10^{42}\,\mbox{erg\,s}^{-1}$, and that therefore does not seem to be solely the result of stellar processes (see the central and right-hand panels of Fig.~\ref{F:Radio_X-ray_lum}). In this waveband, lenticular galaxies split quite evenly among the BPT types, something that does not happen with the PCA ones, where the AC class is again the most populated by far. Nevertheless, unlike the radio domain and despite poorer statistics, the relative frequency distribution of X-ray luminosities for this latter class shows obvious signs of bimodality arising from stark differences between the distributions corresponding to its major BPT contributors, SFs and Seyferts, which peak, respectively, around $10^{41}$ and $10^{42}\,\mbox{erg\,s}^{-1}$, whereas S0--LINERs exhibit intermediate values. This dichotomy means that, irrespectively of the possibility that a substantial fraction of S0--Seyferts may reside in hosts with healthy star formation, the latter is totally insufficient to account for their X-ray emission.

\subsection{Nature of the radio and X-ray emissions}
\label{SS:radio_X-ray_nature}

In this section, we take a closer look at the source of the integrated radio and X-ray emissions of present-day lenticular galaxies. We have restricted our analysis exclusively to objects that belong to the main BPT spectral classes, since this type of classification schemes allows one to differentiate more clearly between AGN and star-formation related activity than broad-band diagnostics.

\begin{figure}
	\includegraphics[width=\columnwidth, viewport=15 1 450 365, clip=true]{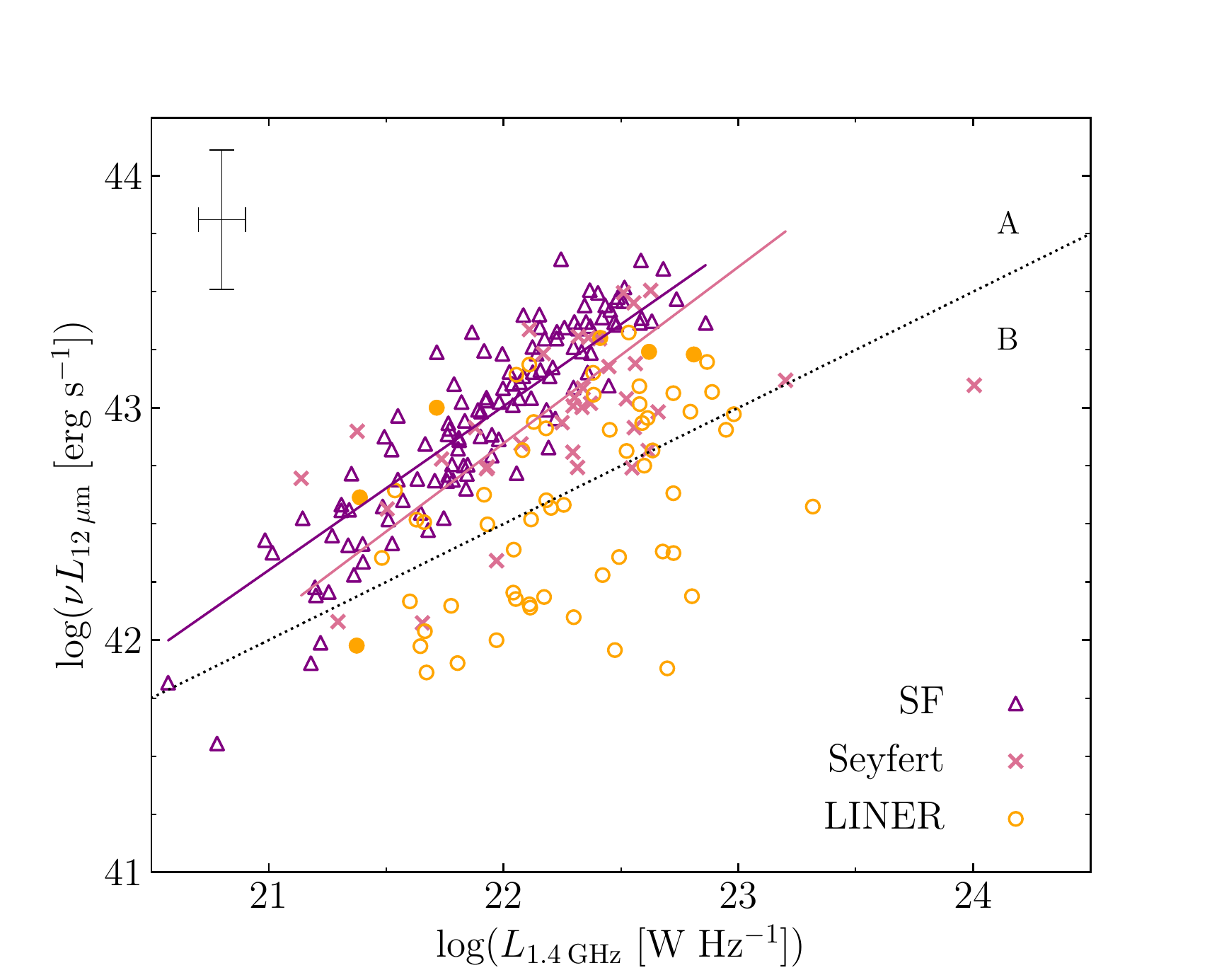}
	\vspace*{-0.5\baselineskip}
    \caption{Observed mid-IR $12\,\mic$ luminosity vs.\ monochromatic 1.4 GHz radio luminosity for the main BPT types of nearby S0 galaxies, SF (purple empty triangles), Seyfert (pink crosses) and LINER (orange empty circles), simultaneously detected in the SDSS, \emph{WISE} and FIRST surveys. Filled symbols are used for the six LINERs that are also X-ray emitters. The dotted line, which divides the diagram in two sectors (letters A and B), is the same as in fig.~3 of \citet{Ros+13}. Also following these authors, we have used an ordinary-least-squares bisector algorithm to estimate the linear regression lines for the SF (purple line) and Seyferts on sector A (pink line). These lines are included in the plot for the sole purpose of allowing the mean trends of both subsets of objects to be compared at a glance. The error bars in the upper-left corner show the typical $1\sigma$ uncertainties of the data for both axes.}
    \label{F:Mid-IR_Radio_lum}
\end{figure}

We begin by examining the distribution of the principal BPT types of our S0 galaxies in the plane of mid-IR and radio luminosities. We have applied the same strategy as \citet{Ros+13}, plotting in Fig.~\ref{F:Mid-IR_Radio_lum} the luminosity at $12~\mic$ derived from the \emph{WISE} \wiii\ band against the monochromatic 1.4 GHz radio power. Following these authors, we also add an eyeball straight line with positive slope across the middle of the diagram to split it into two branches or sectors, A and B, located, respectively, above and below this divider. At this stage, it is no longer surprising to find out that our subset of S0s essentially reproduces the same basic patterns found by \citeauthor{Ros+13} from a considerably larger data set of narrow-line galaxies with diverse morphologies in the redshift range $0.02 < z < 0.15$ located above the \citetalias{Kew+01} curve of the BPT--NII diagram. Thus, we see that the S0--Seyfert and LINER show substantially different mid-IR-to-radio properties, with the former clustering almost completely on sector A and the latter being quite evenly distributed over the diagram -- except those detected in X-rays, which are mostly located in sector A. In this context, it has to be noted that the location of our \Ha-weak and presumably not very dusty S0--LINER in this subspace appears difficult to explain by higher-than-average levels of AGN-heated dust in their hosts or shocks from radio jets. This makes it more plausible to think of an explanation for the radio emission of these objects linked to the diffuse UV field from post-AGB stars, although, as already stated in section~\ref{SS:WHAN}, this scenario might not apply to those LINERs with relatively high values of the \Nii/\Ha\ flux ratio and/or that are X-ray emitters \citetext{\citealt{GonM+09,Caz+18}; see also Fig.~\ref{F:Mid-IR_Radio_lum}}, which are likely powered by AGNs. For their part, the members of the SF class also lie almost exclusively on sector A, generally overlapping with the Seyferts but showing a higher coherence, as well as indications of a somewhat stronger mid-IR emission for a given radio luminosity. Although the variance in the mid-IR-to-radio properties is not statistically significant and, therefore, inconclusive regarding the connection that may exist between the radio emission of Seyferts and the SFR of their hosts, the addition of a second dimension to the radio photometry hints at the possibility that both factors may not necessarily be closely related in local S0 galaxies. 

\begin{figure}
	\includegraphics[width=\columnwidth, viewport=15 -5 400 645, clip=true]{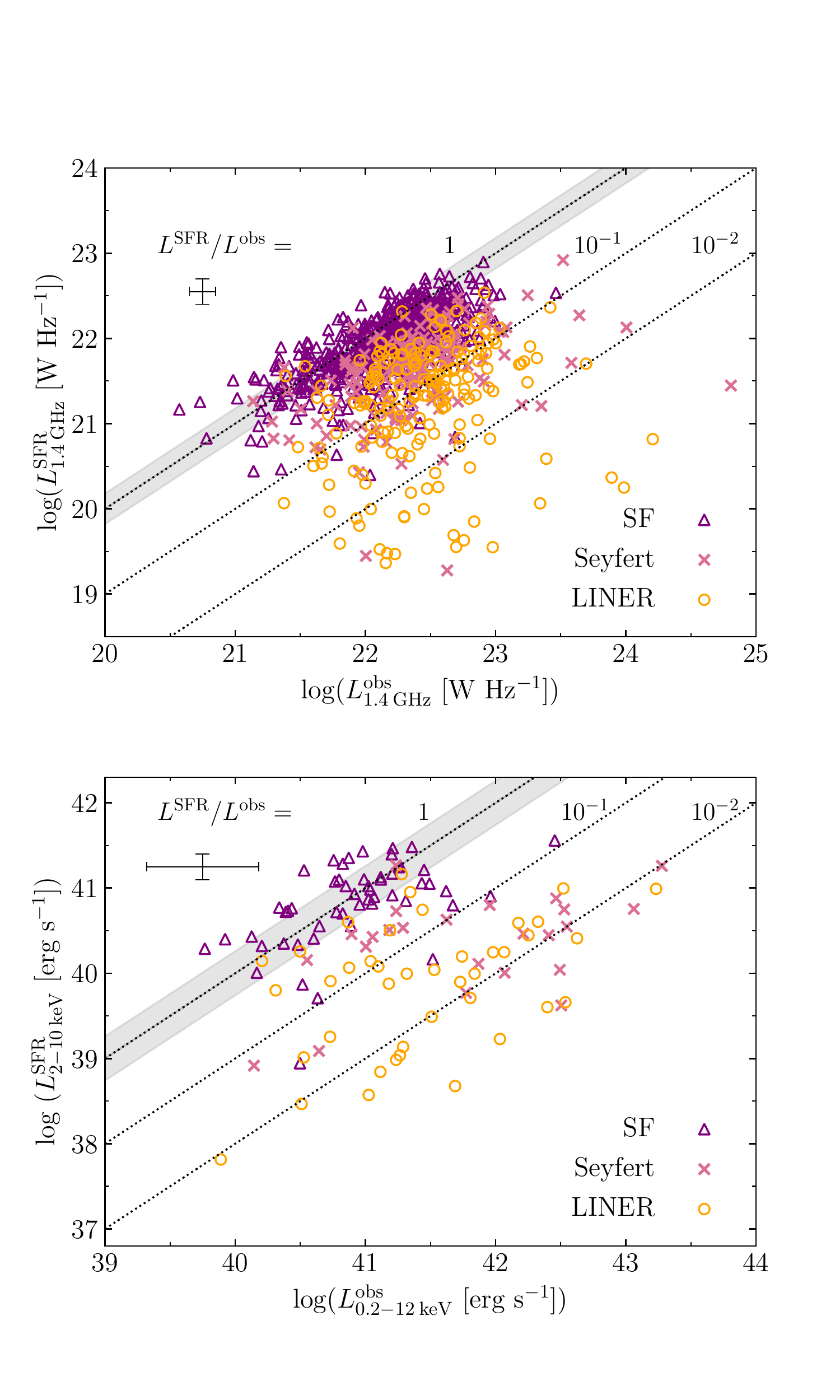}
	\vspace*{-3.5\baselineskip}
    \caption{Monochromatic $1.4$~GHz radio (\emph{top}) and hard-band (2--10~keV) X-ray  luminosities (\emph{bottom}) predicted from the measured SFR, $L^\mathrm{SFR}$, vs.\ the observed values of these quantities, $L^\mathrm{obs}$ (in X-rays the broad-band $L_{\mathrm{0.2-12\;keV}}$ is used as a proxy), for the main BPT types of nearby S0 galaxies. The SFR-calibrated radio and X-ray powers are evaluated using equations (\ref{sfr-radio}) and (\ref{sfr-Xray}), respectively. The dotted lines in both panels correspond to $L^\mathrm{SFR}/L^\mathrm{obs}$ of 1, $10^{-1}$ and $10^{-2}$. The error bars displayed below the $L^\mathrm{SFR}/L^\mathrm{obs}$ legends show the typical $1\sigma$ uncertainties of the data in both panels. The $1\sigma$ uncertainties of the $L^\mathrm{SFR}$ estimates are also represented by the semi transparent grey bands drawn on top of the $L^\mathrm{SFR}/L^\mathrm{obs}=1$ lines. The colour and shape coding of the data points are the same as in Fig.~\ref{F:Mid-IR_Radio_lum}.}
    \label{F:Radio_X-ray_SFR}
\end{figure}

The direct calibration of the expected radio and X-ray fluxes in terms of the measured SFR confirms these suspicions. In Fig.~\ref{F:Radio_X-ray_SFR}, we plot for the subsets of SF, Seyfert and LINER lenticulars, the predicted monochromatic $1.4$~GHz radio (top panel) and hard (2--10~keV) X-ray power (bottom panel), evaluated by replacing, respectively, in equations~(\ref{sfr-radio}) and (\ref{sfr-Xray}) the SFR reported in the GSWLC-2 catalog -- and derived from UV-to-mid-IR SED fitting --, against the observed values of these quantities. Although these graphs are only exploratory due to the smallness and more than likely incompleteness of the samples involved, they reveal a clear displacement between SF and both Seyfert and LINER sources by about one to two orders of magnitude at a given observed luminosity. Thus, while the measured luminosity in both bands for most SF lenticulars is consistent within the uncertainties with being essentially powered by the formation of new stars, the main contribution to the radio-continuum and X-ray emission for most of their counterparts classified as Seyfert and LINER is unrelated to star formation.

This turns out to be particularly true in the X-ray domain (bottom panel of Fig.~\ref{F:Radio_X-ray_SFR}), where the mean of $\log(L^\mathrm{SFR}/L^\mathrm{obs})$ for the subset of SF lenticulars is $-0.089\pm 0.13$ dex ($95\%$ confidence interval) and the standard deviation $0.48$ dex. In contrast, for the subsets of S0--Seyferts and LINERs the means are $-1.4\pm 0.32$ and $-1.6\pm 0.25$ dex, and the standard deviations $0.75$ and $0.79$ dex, respectively, which points to a generally very limited contribution of the SFR to the power radiated in this band by these two spectral classes. Note that only the 4XMM-DR11 data are shown in the plot, where we have approximated $L^\mathrm{obs}$ by the broad-band 0.2--12~keV luminosities, since there are no hard-band 2--10~keV values listed in this catalog. None the less, we have verified that restricting $L^\mathrm{obs}$ to the hardest band (4.5--12 keV) used in the 4XMM-DR11 data processing or approximating it by the total 0.5--7 keV band flux measured by \emph{Chandra}'s instruments does not change our conclusions. It seems, therefore, that the intrinsic absorption and the possible thermal emission from gravitationally heated hot halo gas that can distort the SF-related X-ray emission at energies below $\lesssim 2$~keV do not introduce significant differences in the outcome of this diagnostic. This is in agreement with the finding by \citet{RCS03} that for local star-forming galaxies the relationships between the SFR and both the hard and soft X-ray luminosities obey, in a log-log plot, linear laws with only a slight difference in the intercept.

The relationship between predicted and observed luminosities at 1.4 GHz for local S0s shown in the top panel of Fig.~\ref{F:Radio_X-ray_SFR} follows a pattern very similar to that outlined by the X-ray data. Note, for example, that much as with the X-ray measurements, there are quite a few lenticulars classified as LINER in the upper panel of Fig.~\ref{F:Radio_X-ray_SFR} for which the fraction of the radio luminosity attributable to the birth of new stars does not exceed one per cent at best. In this band, the estimates of the mean values, their $95\%$ confidence limits and the standard deviations of the random variable $\log(L^\mathrm{SFR}/L^\mathrm{obs})$ inferred from the subsets of SF, Seyfert and LINER lenticulars are, respectively, $-0.21\pm 0.025$ and $0.31$ dex, $-0.77\pm 0.085$ and $0.55$ dex, and $-1.2\pm 0.11$ and $0.77$ dex (for these calculations we have used a linear regression model with errors in both variables).

This same test can also be easily reconverted into a diagnostic for potentially misclassified sources, by simply adopting the condition $\log(L^\mathrm{SFR}/L^\mathrm{obs}) > -0.3$ as the definition of an SF galaxy. Thus, by accounting for the uncertainties in both axes, the estimated contamination of the population of SF galaxies by sources whose observed fluxes are not actually related to stellar birth (Seyfert and LINER) is about 8 and 6 per cent in the radio and X-ray bands, respectively ($\sim 9$ per cent for the X-rays had we used for $L^\mathrm{obs}$ the luminosity in the 4.5--12 keV band instead). Conversely, about two per cent of the radio fluxes from Seyfert and LINER sources could actually be explained by looking at their SFR, while according to the X-ray data there is a zero probability that members of these two spectral classes have been given an erroneous optical spectroscopic taxonomy. Although admittedly crude, these estimates serve to rule out that the conclusions of this work may suffer from misclassification issues.

\section{Summary}
\label{S:summary}

We have performed a comprehensive statistical study of the activity, both linked to normal stellar processes as to black hole accretion, of the S0 population in the local Universe ($z\lesssim 0.1$) on an unprecedented scale, not only with regard to the size of our sample of lenticular galaxies, the largest of its type thus far, but also by the number of energy ranges (optical, mid-IR, radio, and X-ray), and the typology (spectral and photometric) of measurements involved.

Building on the substantial data from the NASA-Sloan Atlas, we have first extensively reviewed a number of classification schemes used in the literature to divide our sample of S0 galaxies into physically motivated activity classes according to diagnostics based on specific features of their optical spectra. More specifically, we have characterised the degree and nature of activity of present-day S0 galaxies using the classification paradigm recently introduced by \citet{TSP20}, which is grounded on the PCA of the entire optical spectra, as well as the well-known BPT and WHAN diagrams, which rely on the strength of narrow emission lines. Additionally, we have harnessed the resemblance between the PCA and WHAN dividers to introduce a new activity diagnostic that combines both schemes. We have also carried out an intercomparison of various PDF from the objects belonging to the different spectral categories in order to characterise their properties (stellar mass, redshift) and gain insight on the role played by external factors (local density of neighbours) in determining membership to the different spectral classes. All this has been followed by an exploration of the origin of the mid-IR emission of the main BPT/WHAN and PCA classes in which our S0s are divided, using the photometric data from the \emph{WISE} all-sky survey to further our knowledge on the connection between the spectroscopically-defined taxonomy and alternative, non-spectroscopic classification schemes, such as mid-IR colour--colour diagrams. Last but not least, we have used the FIRST 1.4 GHz photometry in combination with broad-band flux measurements from the \emph{XMM--Newton} and \emph{Chandra} observatories to investigate the interplay between the radio and X-ray luminosities of our S0 galaxies, characterise them according to the spectral type, and apply some straightforward empirical tests for the purpose of unveiling the relationship between the power emitted in these extreme wavebands and star formation.

We summarize below the main results obtained:
\begin{enumerate}
\item The intercomparison of the stellar mass and redshift distributions of the various spectral classes of present-day S0 galaxies has evidenced that those with ongoing star formation tend to be less massive than the rest of their (both active and passive) counterparts, mimicking the behaviour observed in the general galaxy population. This manifestation of downsizing in the local Universe suggests that the imprint of this phenomenon could also be preserved among individual Hubble types. 
\item We confirm previous findings that nearby S0 galaxies showing an active optical spectrum, particularly if they belong to the SF class, display a clear preference for residing in sparser environments of the local Universe, while objects with a fully passive character tend to be more abundant in crowded regions.
\item We have revealed the existence of strong parallels between the demarcations along the vertical axis of the WHAN diagram, based on the EW of the \Ha\ line, and the PS/TR/AC boundaries inferred from the PCA of the entire optical spectrum recently introduced in the first paper of this series to classify S0 galaxies. The coincidence between dividers that depend on such different spectral elements emanates from the tight underlying correlation between the strength of the \Ha, and other prominent nebular emission lines, and the shape of the continuum of the stellar component. This has led us to propose the replacement of the EW(\Ha) in this diagram with the less-prone-to-errors distance to the PS ridge in the PC1--PC2 plane, adding robustness to the classification without altering the results.
\item The \emph{WISE} mid-IR colours of our sample of nearby S0s validate their narrow-line and broad-band optical spectroscopic classifications. We have found that \wxxiii\ is closely related to the distance to the MS of galaxies. This, and the substantial differences observed in the spatial distributions of the various spectral classes in \emph{WISE} colour--colour plots, lend support to the idea that different physical mechanisms take part in their classification. The IR emission of local S0s also appears to rule out the existence of a sizeable fraction of obscured AGNs within this population, shows that members of the weakly active LINER and TR classes are preferentially located in the IRTZ, and, for the objects included in the Seyfert class, does not seem to be predominantly driven by star formation.
\item The active members of our S0 sample faithfully replicate the markedly different trends followed in the mid-IR radio plane by LINER galaxies of all morphologies relative to their SF and Seyfert counterparts. The fact that such a strong segregation is maintained by the presumably gas- and dust-poor LINER lenticulars suggests that at least one part of the members of this class is neither illuminated by new stars nor by AGNs. Instead, the emission lines of these objects are probably generated by the diffuse UV field of evolved stars, as proposed by \citet{CidF+11} and \citet{Sin+13}, among others.
\item Several lines of evidence, among them, the calibration of the radio and X-ray integral luminosities of the active members of our S0 sample associated with their global SFRs published in the literature, indicate that in most of the current S0 classified as LINER, but also in much of those belonging to the Seyfert class, star birth is not driving their emission in these extreme wavebands. This finding seems to contradict previous claims \citep[][]{Bon+15} that radio luminosity in radio-quiet AGNs is mainly due to star-forming-heated emission, and therefore that the radio power is also a good tracer for the SFR of this type of galaxies. We note, however, that in this particular work the galaxy sample used is very different from ours, as it is dominated by systems with $z\sim 1$--2, a time when the volume-averaged cosmic SFR was at its peak \citep[e.g.][]{MD14}. Thus, the discrepancy could simply mean that such an  assertion cannot be extrapolated to galaxies in the local Universe.
\end{enumerate}

Further confirmation of all these results may come from multiwavelength studies working with a single galaxy sample made up of objects detected simultaneously in different energy ranges. This is something utterly impossible to do with the current data, since after applying to our baseline S0 data set all the quality filters established for each one of the four wavebands investigated, we are left with less than 10 sources verifying all constraints, a number totally insufficient to draw any valid conclusion. Finally, we want to highlight that the significant overlap between the spectral properties of current S0 galaxies and those of earlier and later morphologies, as well as the remarkable similarities that, as we have been emphasizing, show their respective spatial distributions in the different types of diagrams analysed, suggest that many of the findings of this work are exportable to other Hubble types.

\section*{Acknowledgements}

We wish to thank an anonymous referee for his/her thorough review of the manuscript, which has helped us improve the presentation of the contents and better emphasize the results. We acknowledge financial support from the Spanish state agency MCIN/AEI/10.13039/501100011033 and by `ERDF A way of making Europe' funds through research grants PID2019--106027GB--C41 and PID2019--106027GB--C43. MCIN/AEI/10.13039/501100011033 has also provided additional support through the Centre of Excellence Severo Ochoa's award for the Instituto de Astrof\'\i sica de Andaluc\'\i a under contract SEV--2017--0709 and the Centre of Excellence Mar\'\i a de Maeztu's award for the Institut de Ci\`encies del Cosmos at the Universitat de Barcelona under contract CEX2019--000918--M. JLT acknowledges financial support by the PRE2020--091838 grant from MCIN/AEI/10.13039/501100011033 and by `FSE Invests in your future'. We are also grateful to all the people involved in the gathering, reduction, and processing of the data sets listed in the Data Availability Statement below, as well as the public and private institutions that have provided the necessary funding, resources and technical support to make possible both the relevant surveys and the release of their measurements to the community. We equally acknowledge the use of the \textsc{TOPCAT} software package (written by Mark Taylor, University of Bristol) in the testing of the data from the various catalogs.

\section*{Data availability}

This research has made prominent use of the following databases in the public domain: the NASA-Sloan Atlas at \url{http://nsatlas.org/}, the Morphological catalog for SDSS galaxies at  \url{https://cdsarc.cds.unistra.fr/viz-bin/cat/J/MNRAS/476/3661}, the Portsmouth Stellar Kinematics and Emission Line Fluxes at \url{https://data.sdss.org/sas/dr12/boss/spectro/redux/galaxy/v1_1/}, the AllWISE Data Release at \url{https://wise2.ipac.caltech.edu/docs/release/allwise/}, the FIRST Survey Catalog at \url{http://sundog.stsci.edu/first/catalogs/readme.html}, the 4XMM--DR11 at  \url{http://xmmssc.irap.omp.eu/Catalogue/4XMM-DR11/4XMM_DR11.html}, as well as the Chandra Source Catalog Release 2.0 at \url{https://cxc.cfa.harvard.edu/csc/}.




\bibliographystyle{mnras}
\bibliography{biblio2}



\onecolumn
\appendix
\newpage

\section{Mean optical spectrum and main eigenspectra from nearby S0 galaxies}
\label{A:eigenspectra}
A data set of $N$ SDSS galaxy spectra can be thought of as a set of flux vectors, $\mathbold{f}_i$, in a space of $\sim 3800$ dimensions or (vacuum) wavelength intervals. Each one of these vectors can be expressed exactly as the sum of a vector of mean fluxes, $\langle\mathbold{f}\rangle$, and a linear combination of $M$ orthonormal vectors or eigenspectra, $\mathbold{e}_j$, that account for all the variability around the mean, in the form \citep[e.g.][]{Con+95}:
\begin{equation}
    \mathbold{f}_i = \langle\mathbold{f}\rangle + \sum_{j=1}^{M} \mbox{PC}j_i\ \mathbold{e}_j\ \ \ \ \ \ \ \ \ \ \ 1\leq i\le N\;,
    \label{eq:flux_approximation}
\end{equation}
where the coefficients PC$j_i\equiv \mathbold{f}_i\cdot \mathbold{e}_j$ are the projections of each individual spectrum on the new base of principal components, which are arranged in decreasing order of their relative power, i.e., of the amount of the total variance of the data that they explain. Thus, when the main variance of a data set lies in a low-dimensional space, one can get a good visualization of it by truncating the above expansion to the first few eigenvectors. In the present case, we showed in \citetalias{TSP20} that near 90 per cent of the sample variance lies in the 2D subspace where we carry out our spectral classification and whose axes are the first two eigenspectra. This is a huge reduction in data size, with a modest loss of information. For those interested in applying this sort of dimensionality reduction to the SDSS galaxy spectra, we provide in Table~\ref{T:components} the mean spectrum and first five eigenspectra inferred from our sample of processed\footnote{As detailed in Sec.\ 3 of \citetalias{TSP20}, to remove any extrinsic source of variability it is necessary to put all the original spectra on an equal basis while preserving their shapes by shifting them to the laboratory rest-frame, and re-binning and normalizing their fluxes.} extinction-corrected SDSS spectra of nearby S0 galaxies, with which one can explain more than 96 per cent of the variation in the sample.

\begin{table}
\centering
\caption{Mean optical spectrum and first five eigenspectra inferred from S0 galaxies with $z\leq 0.1$.}
\label{T:components}
\begin{tabular}{ccccccc} 
\hline
		Wavelength [\AA] & Mean flux & First eigenspectum & Second eigenspectrum & Third eigenspectrum & Fourth eigenspectrum & Fifth eigenspectrum\\
		\hline
3900.3 & 0.54156 &  0.02360 & -0.00375 & -0.01112 & -0.02274 & -0.03610\\
3901.2 & 0.54442 &  0.02451 & -0.00393 & -0.01197 & -0.02397 & -0.03791\\
3902.1 & 0.54509  & 0.02537 & -0.00439 & -0.01228 & -0.02550 & -0.04051\\
3903.0 & 0.54257 &  0.02606 & -0.00448 & -0.01267 & -0.02605 & -0.04073\\
3903.9 & 0.53638 &  0.02635 & -0.00409 & -0.01282 & -0.02631 & -0.04276\\
\hline
\multicolumn{7}{l}{This is a sample table consisting of the first 5 rows of data. A machine-readable version of the full table is available online at the CDS website.}
\end{tabular}
\end{table}

\bsp	
\label{lastpage}
\end{document}